\documentclass[aps,prl,reprint,groupedaddress,superscriptaddress]{revtex4-2}
\usepackage[T1]{fontenc}
\usepackage{graphicx}
\usepackage{dcolumn}
\usepackage{bm}
\usepackage{dsfont}
\usepackage{amssymb}  
\usepackage{amsmath}
\usepackage{lmodern}
\usepackage{mathtools}
\usepackage{mathrsfs}
\usepackage{textgreek}
\usepackage{xr-hyper}
\usepackage[colorlinks,citecolor=black,linkcolor=black,urlcolor=black]{hyperref}
\usepackage{ragged2e}
\usepackage[normalem]{ulem}
\usepackage{upgreek} 
\usepackage{siunitx}
\usepackage{xcolor}

\usepackage{lineno}
\usepackage{soul}
\begin{document}
\title[]{Electrically tunable momentum space polarization singularities in liquid crystal microcavities}

\author{Przemys\l{}aw\,Oliwa}
\affiliation{Institute of Experimental Physics, Faculty of Physics, University of Warsaw, ul.~Pasteura 5, PL-02-093 Warsaw, Poland}
\author{Piotr Kapu\'sci\'nski}
\affiliation{Institute of Experimental Physics, Faculty of Physics, University of Warsaw, ul.~Pasteura 5, PL-02-093 Warsaw, Poland}
\author{Maria Pop\l{}awska}
\affiliation{Institute of Experimental Physics, Faculty of Physics, University of Warsaw, ul.~Pasteura 5, PL-02-093 Warsaw, Poland}
\author{Marcin Muszy\'nski}
\affiliation{Institute of Experimental Physics, Faculty of Physics, University of Warsaw, ul.~Pasteura 5, PL-02-093 Warsaw, Poland}
\author{Mateusz Kr\'ol}
\affiliation{Institute of Experimental Physics, Faculty of Physics, University of Warsaw, ul.~Pasteura 5, PL-02-093 Warsaw, Poland}
\author{Przemys\l{}aw Morawiak}
\affiliation{Institue of Applied Physics, Military University of Technology, ul.~gen.~Kaliskiego 2, PL-00-908 Warsaw, Poland}
\author{Rafa\l{} Mazur}
\affiliation{Institue of Applied Physics, Military University of Technology, ul.~gen.~Kaliskiego 2, PL-00-908 Warsaw, Poland}
\author{Wiktor Piecek}
\affiliation{Institue of Applied Physics, Military University of Technology, ul.~gen.~Kaliskiego 2, PL-00-908 Warsaw, Poland}
\author{Przemys\l{}aw Kula}
\affiliation{Institue of Chemistry, Military University of Technology, ul.~gen.~Kaliskiego 2, PL-00-908 Warsaw, Poland}
\author{Witold Bardyszewski}
\affiliation{Institute of Theoretical Physics, Faculty of Physics, University of Warsaw, ul.~Pasteura 5, PL-02-093 Warsaw, Poland}
\author{Barbara Pi\k{e}tka}
\affiliation{Institute of Experimental Physics, Faculty of Physics, University of Warsaw, ul.~Pasteura 5, PL-02-093 Warsaw, Poland}
\author{Helgi Sigur{\dh}sson}
\email{Helgi.Sigurdsson@fuw.edu.pl}
\affiliation{Institute of Experimental Physics, Faculty of Physics, University of Warsaw, ul.~Pasteura 5, PL-02-093 Warsaw, Poland}
\affiliation{Science Institute, University of Iceland, Dunhagi 3, IS-107, Reykjavik, Iceland}
\author{Jacek Szczytko}
\email{Jacek.Szczytko@fuw.edu.pl}
\affiliation{Institute of Experimental Physics, Faculty of Physics, University of Warsaw, ul.~Pasteura 5, PL-02-093 Warsaw, Poland}

\begin{abstract}
    Momentum space polarization singularities of light appear as vectorial twists in the scattered and radiated far field patterns of exotic photonic structures. They relate to important concepts such as bound states in the continuum, spatiotemporal light steering, polarization M\"obius strips, Berry curvature and associated topological photonic phenomena. Polarization singularities, such as completely circularly polarized C-points, are readily designed in real space through interference of differently polarized beams. In momentum space, they require instead sophisticated patterning of photonic crystal slabs of reduced symmetries in order to appear in the corresponding band structure with scarce in-situ tunability. Here, we show that momentum space singularities can be generated and, importantly, electrically tuned in the band structure of a highly birefringent planar liquid crystal microcavity that retains many symmetries. Our results agree with theoretical predictions and offer exciting possibilities for integration of momentum space polarization singularities in spinoptronic technologies.
\end{abstract}

\maketitle
\section{Introduction}
Polarization singularities known as C-points are twisted sources of highly circularly polarized light encircled by a quantized winding linear polarization vector in the electromagnetic field's parameter space~\cite{Ruchi_Hindawi2020, Wang_APL2021}. More formally, they are phase-singular points in the electromagnetic scalar field $\bm{E} \cdot \bm{E}$ where the orientation of the polarization ellipse is undefined. They possess topological charges intimately related to the field's geometric (Berry) phase~\cite{Bliokh_IOP2019} and share strong similarities with electromagnetic textures known as optical skyrmions~\cite{Shen_NatPhotRev2024}. C-points can be readily generated in real space through superposition of beams with orthogonal photon spin- and orbital angular momentum~\cite{Ruchi_Hindawi2020} utilizing e.g. q-plates based on metasurfaces and/or liquid crystals technologies~\cite{Milione_PRL2011, Cardano_OptExpr2013}. Only, recently has the interest of researchers been piqued towards polarization singularities in momentum (reciprocal) space~\cite{Zeng_PRL2021, Zhang_PRL2018} due to their implication for polarization generation~\cite{Guo_AdvOptMat2019}, routing chiral emission in atomic monolayers~\cite{Wang_LScAppl2020}, laser transmission through birefringent liquid crystal layers \cite{Kiselev_(2007)_J.Phys.Condens.Matter},polarimetry mapping of photonic Berry curvature~\cite{Guo_PRL2020}, the far field radiation properties of patterned photonic structures hosting bound states in continuum (BICs)~\cite{Doeleman_NatPhot2018, Liu_PRL2019, Yin_Nature2020, Wang_FrontPhys2022, Ye_PRL2020, Qin_LSA2023} or creation of polarization M\"obius strips \cite{Bliokh_(2021)_Phys.Fluids}, etc. 

Similar to scalar phase singularities like in optical vortex beams~\cite{Shen_LSAppl2019} polarization singularities in multicomponent fields can be assigned a topological charge referred to as {\it Poincar\'{e}-Hopf index} defined by a closed path integral in the field's parameter subspace,
\begin{equation}
2w = \frac{1}{2 \pi} \oint \nabla \phi_{\mu,\nu} \cdot \mathrm{d} \bm{l} = 1,2,3,\dots
\end{equation}
The factor 2 is conventional and relates C-points to the winding numbers of optical merons or half-charge skyrmions~\cite{Shen_NatPhotRev2024}. Here, $\phi_{\mu,\nu} = \arg{\left(s_\mu + i s_\nu \right)}$ is the {\it Stokes phase}~\cite{Bliokh_IOP2019} containing the singularity and $s_{i} = \bm{E}^\dagger \cdot \bm{\sigma}_{i} \bm{E} / \left|\bm{E} \right|^2$ are the Stokes parameters and $\bm{E} = \left[E_{H}, E_{V} \right]^T$ is the paraxial cavity field in the linear polarization basis, and $\bm{\sigma} = \left(\bm{\sigma}_{z}, \bm{\sigma}_{x}, \bm{\sigma}_{y}\right)$ are the Pauli matrices. 
In contrast to typical scalar vortices which have cores of zero intensity, the vectorial nature of the electromagnetic field permits C-points to be finite and fully circularly polarized at their core ($s_3 = \pm1$) -- hence the name -- but of undefined linear polarization. There also exist nonradiative V-points which form points of zero total field intensity connected to BICs~\cite{Wang_FrontPhys2022}, and L-lines, trajectories of pure linear polarization which separate regions of opposite handedness belonging to C-points~\cite{Freund_OptLett2002}.

With surging advancement in both beam steering and polarization manipulation within the nanophotonics sector through the use of miniaturised metasurfaces, photonic crystal slabs, and gratings, new opportunities have opened to develop optical technologies based on polarization singularities in the far field reciprocal space~\cite{Wang_FrontPhys2022, Liu_NanoPho2021, Ni_Science2021}. They could become a disruptive force for classical or quantum information processing purposes, enhanced light-matter interaction in optical cavities, optical trapping, chiral sensing and lasing, etc~\cite{Ruchi_Hindawi2020, Wang_APL2021}. This development underpins a need in designing microscale optical systems that not only passively generate pre-determined polarization singularities, but can also be flexibly tuned in-situ to smoothly adjust C-point characteristics and their number. So far, implementations of momentum space polarization singularities rely mostly on irreversibly patterned materials such as photonic~\cite{Doeleman_NatPhot2018, Zeng_PRL2021}, polaritonic~\cite{Kim_NanoLett2021, Dang_AdvOptMat2022, Ardizzone_Nature2022} or plasmonic~\cite{Zhang_PRL2018, Chen_PRB2019} crystal slabs or heterostructures. Control over the singularity type and location in reciprocal space is achieved by reducing or breaking symmetries of the slab such going from four-fold ($C_{4v}$) to two-fold ($C_{2v}$) rotational in-plane symmetry~\cite{Chen_PRB2019}, or the up-down~\cite{Yin_Nature2020} or in-plane~\cite{Liu_PRL2019} inversion symmetry. However, once the sample is created its properties are set and the singularities cannot be tuned or adjusted unless another sample with new parameters is created. This severely limits the application of momentum space polarization singularities in optical technologies and experiments requiring in-situ tuning. 

Here, we overcome this challenge, and demonstrate electrically tunable multiple half-charge momentum space polarization singularities in the transmitted far field of an optical microcavity filled with uniaxial nematic liquid crystal as recently proposed~\cite{Zhao2023}. The singularities stem from the strong birefringence of the anisotropic liquid crystal microcavity (LCMC) which leads to effective photonic spin-orbit coupling (SOC)~\cite{Bliokh_NatPhot2015} and emergent optical activity, despite many cavity symmetries staying intact. We identify two types of singularities, non-degenerate C-points characterized by strong circular polarization, and degenerate diabolical points (DPs) which break into pairs of exceptional points (EPs) through polarization dependent losses \cite{Krol_NatComm2022}.

\section{Results}
\subsection{Liquid crystal microcavity sample}

\begin{figure*}
    \centering
    \includegraphics{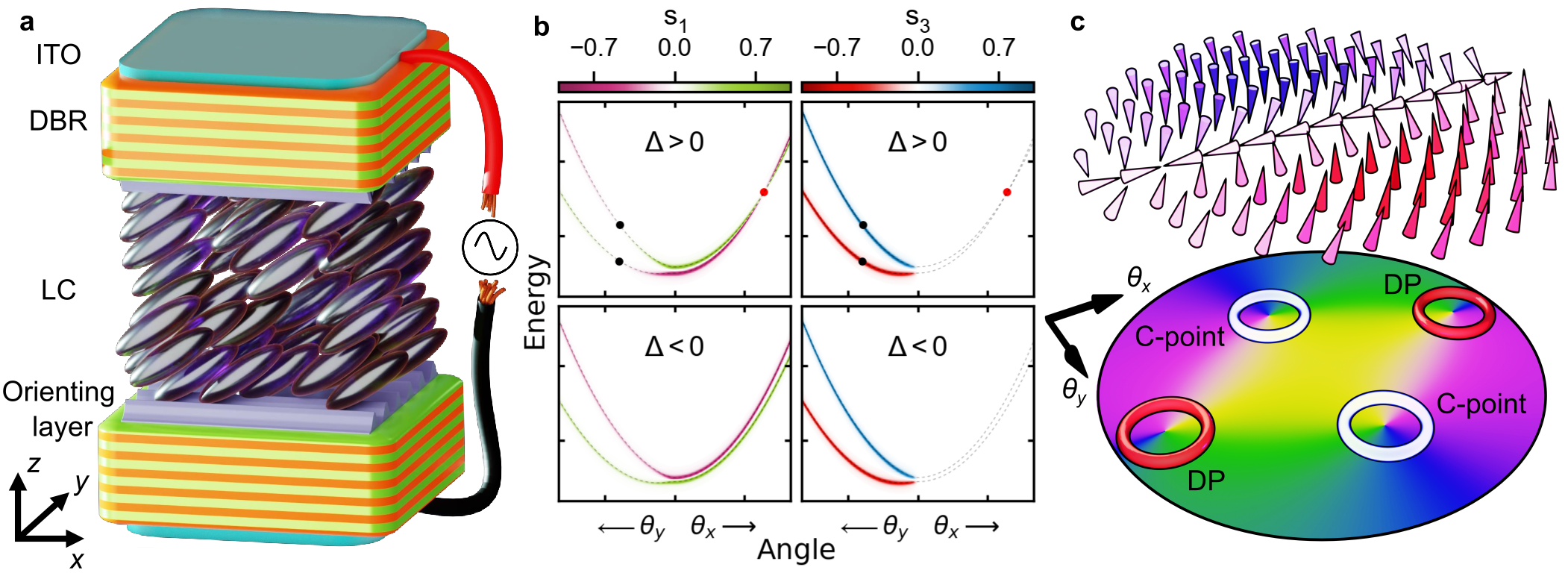}
    \caption{\textbf{Liquid crystal microcavity device alongside the cavity photon dispersion and associated momentum space polarization singularities.}  \textbf{a} Schematic figure of the electrically tunable LCMC device. \textbf{b} Calculated polarization resolved cavity dispersion relation around the equal Rashba-Dresselhaus SOC regime for two different Stokes parameters $s_{1,3}$ and detunings $\Delta$. The black and red dots denote the location of C-points and diabolical points, respectively. \textbf{c} Schematic illustration of the Stokes vector field showing two $w=1/2$ lemon-type C-points and two $w=-1/2$ star-type diabolical points.}
    \label{Figure main - Idea figure}
\end{figure*}

The studied LCMC is presented schematically in Fig.~\ref{Figure main - Idea figure}\textbf{a}. It consists of a birefringent medium layer -- uniaxial liquid crystal (LC) in a nematic phase -- embedded between two parallel distributed Bragg reflectors (DBRs). The transparent ITO electrodes covered by the two DBRs allow for electrical control over the spatial orientation of the LC molecular director, and therefore over the anisotropy of the refractive index~\cite{Rechcinska_Science2019}. This enables tuning of the energy of the cavity mode (branch) that is polarized in the rotation plane of the molecules, while the mode with polarization orthogonal to this plane remains unaffected (see Figure~\ref{Figure main - Idea figure}a). For simplicity, we refer to the former as \emph{horizontal} (H) polarization and the latter \emph{vertical} (V) polarization, coinciding with the chosen cavity in-plane directions, $\hat{\mathbf{x}}$ and $\hat{\mathbf{y}}$, respectively. 

We work in the Rashba-Dresselhaus (RD) regime~\cite{Rechcinska_Science2019, Krol_PRL2021} where degenerate H and V polarized cavity modes have staggered longitudinal parity. That is, $H \left(m+1\right)$ and $V \left(m\right)$, where $m$ is the mode index, due to the extreme birefringence of the liquid crystal molecules. The cavity in-plane momentum is defined as \mbox{$\bm{k} \equiv \left[k_x, k_y\right]^T$} and is related to the angle of incidence of emitted light through \mbox{$\bm{k} = k_{0} \sin \left(\theta\right) \left[ \cos \left( \varphi \right), \sin \left( \varphi \right) \right]^{T}$}, where $k_{0} = 2\pi/\lambda$, $\theta$ is the angle of incident wave and $\varphi$ is the azimuthal in-plane angle, so we defined $\bm{\theta} = \left[\theta_{x}, \theta_{y} \right]^{T} = \theta\left[\cos\left(\varphi\right), \sin \left(\varphi\right) \right]^{T}$. In what follows, data corresponding to reciprocal space (also referred to as \emph{momentum} space) is plotted as a function of angle to relate directly to the experimental configuration. 

\subsection{Two-mode photonic Hamiltonian}
We start by establishing the presence of the polarization singularities in our RD LCMC system using two different theoretical approaches, both in good agreement with following experimental results. First, we calculate the LCMC dispersion relation using the Berreman and Schubert method \cite{Berreman_(1972)_J.Opt.Soc.Am., Schubert_(1996)_Phys.Rev.B} which has been widely successful including in our past experiments~\cite{Lekenta_LSA2018, Rechcinska_Science2019, Krol_PRL2021, Krol_Optica2021, Lempicka_SciAdv2022, Krol_NatComm2022}. While highly accurate, the method is computationally time-consuming and does not provide much physical insight into the observed polarization singularities. 

For this reason, we also apply more recently developed approach based on optical cavity $\bm{k} \cdot \bm{p}$ perturbation theory~\cite{Oliwa_(2024)_Phys.rev.res.}, offering a simple coupled-mode description in the form of a $2 \times 2$ Hamiltonian. This method provides a more transparent, quantum mechanics inspired, framework to analyze cavity photon SOC effects between two quasiresonant orthogonal linearly-polarized LCMC modes without any phenomenological parameters, based on the Hamiltonian in the form:
\begin{equation}
    \mathcal{H} = h_0 \left(\bm{k} \right) \mathds{1}_{2 \times 2} + \bm{h}(\bm{k}) \cdot \bm{\sigma}.
    \label{equation: 2x2 hamiltonian - simple version}
\end{equation}
The Hamiltonian is written in a rotated polarization basis to properly account for interactions between cavity modes \cite{Oliwa_(2024)_Phys.rev.res.}. Here, $h_i\left(\bm{k}\right)$ are in-plane momentum-dependent coefficients that are complex (rendering this Hamiltonian non-Hermitian) due to the finite-lifetime cavity photon states. We stress that the order of the Pauli matrices is $\bm{\sigma} = \left(\bm{\sigma}_{z}, \bm{\sigma}_{x}, \bm{\sigma}_{y}\right)$. The coefficients depend on the elements of the dielectric tensor~\cite{Supplementary_Materials} and can be written in the parabolic approximation:
\begin{subequations}
    \label{equation: list of h_i parameters}
    \begin{equation}
        h_{0} = \omega_{0} + i \Gamma_{0} + \frac{\hbar k_{x}^{2}}{2m_{x}} + \frac{\hbar k_{y}^{2}}{2m_{y}},
    \end{equation}
    \begin{equation}
        h_{1} = \Delta + i \delta \Gamma + \delta_x k_{x}^{2} + \delta_{y} k_{y}^{2},
    \end{equation}
    \begin{equation}
        h_{2} = 0,
    \end{equation}
    \begin{equation} \label{eq.RD}
        h_{3} = -2\alpha k_{y}.
    \end{equation}
\end{subequations}
Here, $\omega_0 = \mathrm{Re} \left( \omega_{H \left( m + 1 \right)} + \omega_{V \left(m \right)} \right)/2$ and $\Gamma_0 = \mathrm{Im} \left( \omega_{H \left( m + 1 \right)} + \omega_{V \left(m \right)} \right)/2$ are the mutual frequency and decay rate of the H, V cavity photons, and $\Delta = \mathrm{Re} \left( \omega_{H \left( m + 1 \right)} - \omega_{V \left(m \right)} \right)/2$ and $\delta \Gamma = \mathrm{Im} \left( \omega_{H \left( m + 1 \right)} - \omega_{V \left(m \right)} \right)/2$ are their frequency and decay rate differences at normal incidence $\bm{k} = \bm{0}$. The remaining parameters linear and quadratic in $\bm{k}$ are detailed in the SM~\cite{Supplementary_Materials}. 

The eigenvalues and right eigenvectors of~\eqref{equation: 2x2 hamiltonian - simple version} are:
\begin{subequations}
    \begin{equation} \label{eig}
        \omega_{\pm} = h_{0} \pm \sqrt{h_{1}^{2} + h_{3}^{2}},
    \end{equation}
    \begin{equation} \label{eig.states}
        \bm{q}_{\pm} = 
        \begin{bmatrix} 
            h_{1} \pm \sqrt{h_{1}^{2} + h_{3}^{2}} \\
            i h_{3}
        \end{bmatrix}.
    \end{equation}
\end{subequations}
The spectrum of the LCMC is given by the eigenvalues~\eqref{eig} and we refer to $\omega_\pm$ as upper and lower frequency branches.  The eigenstates~\eqref{eig.states} define the Stokes components of the cavity field which, upon transformation to the canonical cavity basis~\cite{kavokin_microcavities_2007}, are written as (not normalized)~\cite{Oliwa_(2024)_Phys.rev.res.}:
\begin{equation} \label{eq.stokes}
\begin{split}   
S_{i, \pm}(\bm{k}) & = \sum_{s, s^{\prime}} q_{\pm, s}^{\ast} q_{\pm, s^{\prime}} \left(\bm{\sigma}_{i, s s^{\prime}} \vphantom{ \sum_{s^{\prime \prime} m^{\prime \prime}}}\right. + \\ 
& \left.\sum_{s^{\prime \prime} m^{\prime \prime}} \mathbf{R}_{s^{\prime \prime} m^{\prime \prime}, s m}^{\ast} \bm{\sigma}_{i, s^{\prime \prime} s^{\prime}} + \bm{\sigma}_{i, s s^{\prime \prime}} \mathbf{R}_{s^{\prime \prime} m^{\prime \prime}, s^{\prime} m^{\prime}}\right)
\end{split}
\end{equation}
where $s$ and $s^{\prime}$ denote vector elements of $\bm{q}_{\pm}$, and $\mathbf{R}_{sm, s^{\prime} m^{\prime}}$ represents an element of the rotation matrix $\mathbf{R}(\bm{k})$ detailed in the SM~\cite{Supplementary_Materials}. The $sm$ and $s^{\prime} m^{\prime}$ elements denote the modes in the Hamiltonian Eq.~\eqref{equation: 2x2 hamiltonian - simple version}, so $sm \in \{H \left( m + 1\right), V \left( m \right)\}$ and the same for $s^{\prime} m^{\prime}$. The $s^{\prime \prime} m^{\prime \prime}$ refers to all the states in the cavity that are different from the $sm$ states. Equation~\eqref{eq.stokes} is cumbersome but, differently from previous 2-mode phenomenological models, it allows accurate {\it ab initio} calculation of the polarization field.

Having established the spectral~\eqref{eig} and polarization measurables~\eqref{eq.stokes} of our system, we can highlight the presence of polarization singularities by plotting the dispersion relation of the $s_{1}$ and $s_3$ Stokes parameters along $k_x$ and $k_y$ in Fig.~\ref{Figure main - Idea figure}\textbf{b} for positive ($\Delta>0$) and negative ($\Delta<0$) detuning. We remind that $s_{1,2,3}$ denote the degree of horizontal-vertical linear polarization,  diagonal-antidiagonal linear polarization, and left- and right-hand circular polarization of the bands, respectively. 

As can be seen in Fig.~\ref{Figure main - Idea figure}\textbf{b}, the bands are strongly circularly polarized along $k_y$ direction (denoted by the blue-red $s_3$ colorscale) due to the Rashba-Dresselhaus term [Eq.~\eqref{eq.RD}]. The singularities can be located in the dispersion for $\Delta > 0$ where the sign of $s_1$ inverts between the bands. Non-degenerate C-points are marked with black dots and a degenerate diabolical point with a red dot. In the former, the degree of linear polarization $\sqrt{s_{1}^2 + s_{2}^2}$ goes to zero while the circular polarization $|s_3|=1$ is maximal. For the latter, the bands cross (in real energy) forming a so-called diabolical point~\cite{Chen_PRB2019} in the Hermitian limit, which breaks into a pair of exceptional points (EPs) in the non-Hermitian case, bridged by a Fermi arc (not shown). In a negatively detuned cavity $\Delta < 0$ (lower panels of Fig.~\ref{Figure main - Idea figure}\textbf{b}) both C-points and diabolical points disappear which we explain in the following.

Figure~\ref{Figure main - Idea figure}\textbf{c} illustrates these singularities more vividly in Stokes vector field and phase $\phi_{1,2}\left(\bm{k}\right)$ of the upper branch. The C-points are identifiable as lemon-type $w=1/2$ polarization singularities whereas the diabolical points as star-type with opposite phase winding $w=-1/2$. The position of the C-points for a given branch can be derived in the Hermitian limit (see SM \cite{Supplementary_Materials}):
\begin{equation}
    \label{equation - main - position of C-points}
    k_{y, CP} = \pm \sqrt{  \frac{\Delta}{ \Sigma\left(m\right) \frac{4 L c \sqrt{m \left(m + 1\right)}}{\pi^{3}\left(2m + 1\right)} \left(\frac{\varepsilon_{xz}}{\varepsilon_{zz}}\right)^{2}  - \delta_{y}}}
\end{equation}
where $L$ is the cavity thickness, $c$ is the speed of light, $\varepsilon_{ij}$ are elements of the dielectric tensor, and $\Sigma \left(m\right)$ describes correction from other cavity branches. Equation~\eqref{equation - main - position of C-points} explains why C-points vanish in negatively detuned microcavity. Although the sign of $\delta_{y}$ varies with $\theta$, the value of $\Sigma$ remains strictly positive and large in the RD regime (see SM \cite{Supplementary_Materials}), ensuring that the denominator is greater than zero. Consequently, within the presented model \mbox{C-points} can only exist when $\Delta > 0$.

For the diabolical points, it is important to stress that, due to non-Hermiticity, they will separate into two pairs of EPs connected by Fermi arcs~\cite{Bergholtz_RMP2021, Chen_PNAS2021, Su_SciAdv2021, Krol_NatComm2022}. However, the EPs are so close to each other at this scale that they appear as half-charge polarization singularities of star-type $w = -1/2$ on opposite sides of the $k_x$ axis. To better observe the structure of the EP pairs, we show zoomed in regions of the results in Figs. S1 and S2 in the SM~\cite{Supplementary_Materials}. We note that an EP would exhibit pure circular polarization, like a C-point, only if the cavity possesses mirror symmetry in the $x-y$ plane \cite{Richter_(2017)_Phys.Rev.A}. However, in this case, the cavity belongs to the $C_{2h}$ symmetry point group with $C_{2}$ axis along $y$ axis, allowing for all Stokes polarization parameters to be non-zero around the EPs.

An EP occurs when $h_{1}^{2} + h_{3}^{2} = 0$, which for $\Delta \gg \delta \Gamma$ yields following equation:
\begin{subequations}
    \label{equation - main - position of EP}
    \begin{equation}
        k_{x, EP} = \pm \sqrt{-\frac{\Delta}{\delta_{x}}}\left(1 + \frac{\delta_{y}\delta \Gamma^{2}}{8 \alpha^{2} \Delta}\right),
    \end{equation}
    \begin{equation}
        k_{y, EP} = \pm \frac{\delta \Gamma}{2\alpha}.
    \end{equation}
\end{subequations}
Hence, we can observe four EPs located near the $k_{x}$ axis. In the Hermitian limit $\delta \Gamma \to 0$ we retrieve the location of the two diabolical points. Equation~\eqref{equation - main - position of EP} provides a condition for the existence of these degeneracy points: $\Delta/\delta_{x} < 0$. Since $\delta_{x} < 0$ for any rotation angle of the molecule $\theta$, the points can only appear in the case of positive detuning $\Delta > 0$. Because the anti-hermitian part of our Hamiltonian is weak (but not negligible) we will---for simplicity---refer to the locality in momentum space hosting EP pairs as a diabolical points (DPs).

\subsection{Experimental results}

\begin{figure*}[t!]
    \centering
    \includegraphics{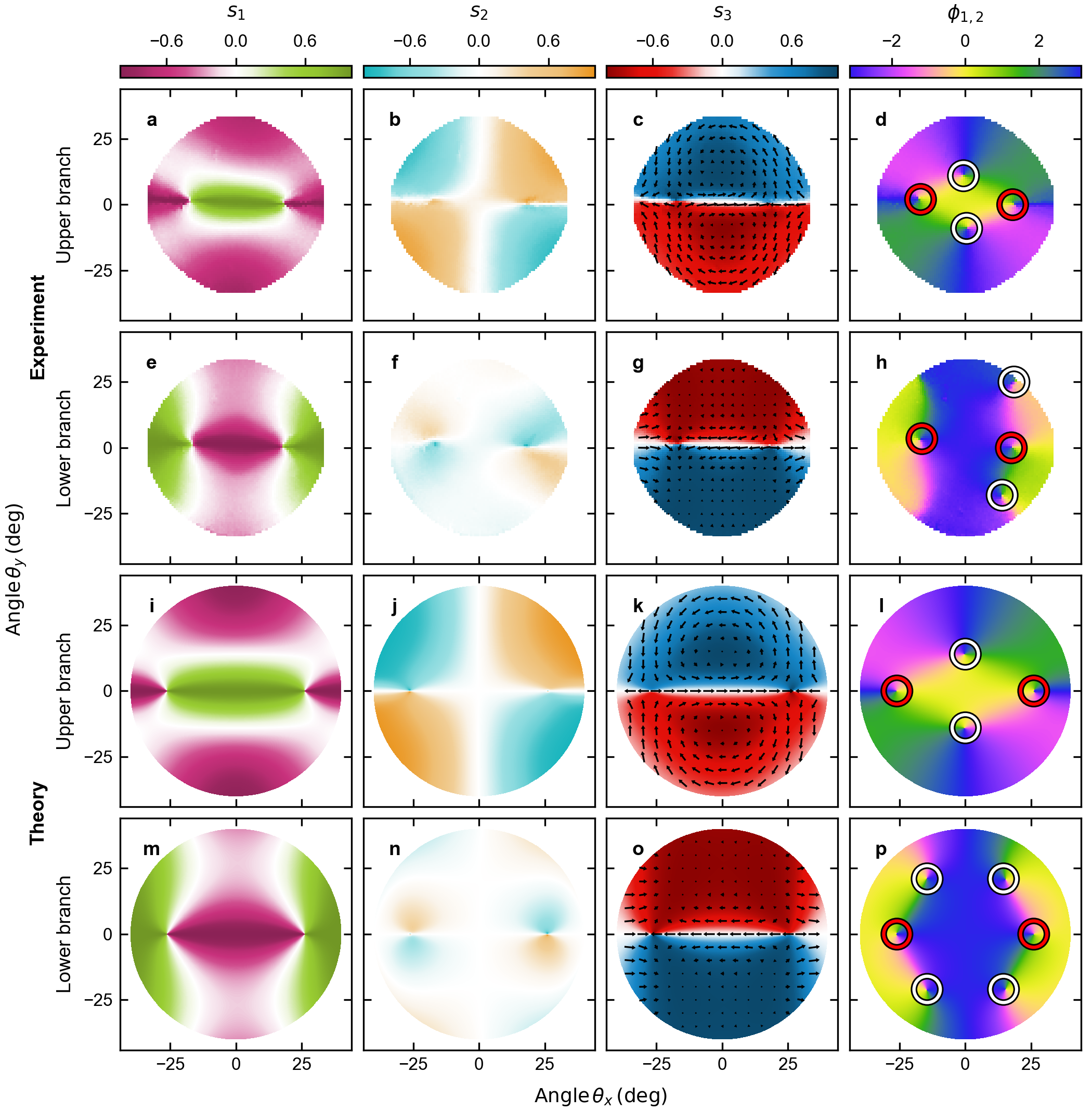}
    \caption{\textbf{Stokes parameters and phase for positive detuning}. \textbf{a-h} Experimentally and \textbf{i-p} theoretically obtained Stokes parameters \textbf{a}, \textbf{e}, \textbf{i}, \textbf{m} $s_1$, \textbf{b}, \textbf{f}, \textbf{j}, \textbf{n}, $s_2$, \textbf{c}, \textbf{g}, \textbf{k}, \textbf{o} $s_3$, and \textbf{d}, \textbf{h}, \textbf{l}, \textbf{p} Stokes phase $\phi_{1,2} = \mathrm{arg}\left(s_1 + i s_2\right)$ for \textbf{a-d, i-l} upper and \textbf{e-h, m-p} lower branch in the case of the positive detuning of $2\hbar \Delta = \hbar \mathrm{Re}\left( \omega_{H \left(m + 1\right)} - \omega_{V \left(m\right)} \right) = 11.0 \,\mathrm{meV}$. The black arrows on the $s_3$ map correspond to $\bm{s}_{\parallel} = \left[s_{1}, s_{2} \right]^{T}$. The white and red circles on the Stokes phase $\phi_{1,2}$ maps mark the position of the C-points and pair of exceptional points, respectively.}
    \label{Figure main - Comparison between theory and experiment - positive detuning}
\end{figure*}

To experimentally verify the presence of polarization singularities in the LCMC dispersion, we performed angle-resolved tomography of polarized white light transmission. We begin with Stokes parameter results for relatively large positive detuning (compared to the mode linewidth of 4 -- 6 meV), $2\hbar\Delta = 11.0 \,\mathrm{meV}$, for the upper branch in Figs.~\ref{Figure main - Comparison between theory and experiment - positive detuning}\textbf{a}-\textbf{d} and the lower branch in \mbox{Figs.~\ref{Figure main - Comparison between theory and experiment - positive detuning}\textbf{e}-\textbf{h}}. The overlaid arrows in Figs.~\ref{Figure main - Comparison between theory and experiment - positive detuning}\textbf{c} and \ref{Figure main - Comparison between theory and experiment - positive detuning}\textbf{g} represent the field's linear polarization, i.e. projection of the Stokes vector $\bm{s}_{\parallel} = \left[s_{1}, s_{2} \right]^{T}$. Figures~\ref{Figure main - Comparison between theory and experiment - positive detuning}\textbf{d} and \ref{Figure main - Comparison between theory and experiment - positive detuning}\textbf{h} show the Stokes phase $\phi_{1,2} = \arg \left(s_{1} + i s_{2} \right)$. 

In agreement with our 2-mode theory [Eq.~\eqref{equation: 2x2 hamiltonian - simple version}] and transfer matrix calculations, the upper branch reveals two lemon-type $w=1/2$ C-points along the $\theta_{x} \approx 0^{\circ}$ direction and two star-type $w=-1/2$ degenerate polarization singularities corresponding to diabolical points along the $\theta_{y} \approx 0^{\circ}$ direction, marked by white and red circles respectively in the Stokes phase map in Fig.~\ref{Figure main - Comparison between theory and experiment - positive detuning}\textbf{d}. The direction of the linear polarization winds around these points resulting in finite curl of the vector field $\bm{s}_{\parallel}$. As expected, near the C-points, we observe a high degree of circular polarization $|s_3| \gg |\bm{s}_{\parallel}|$. In the lower branch, our theory based on transfer matrix calculation reveals four C-points located at \mbox{$\bm{\theta} \approx \left[\pm 14^{\circ}, \pm 21^{\circ} \right]^{T}$} and two DPs at \mbox{$\bm{\theta} \approx \left[\pm 25^{\circ}, 0 \right]^{T}$} (see Figs.~\ref{Figure main - Comparison between theory and experiment - positive detuning}\textbf{m}-\textbf{p}). Since these C-points are located at much higher angles, experimental errors in the measured Stokes parameters can increase, resulting in two C-points escaping just outside our field of view (see Figs.~\ref{Figure main - Comparison between theory and experiment - positive detuning}\textbf{e}-\textbf{h}). None-the-less, the agreement between theory and experiment is striking.

\begin{figure}
    \centering
    \includegraphics{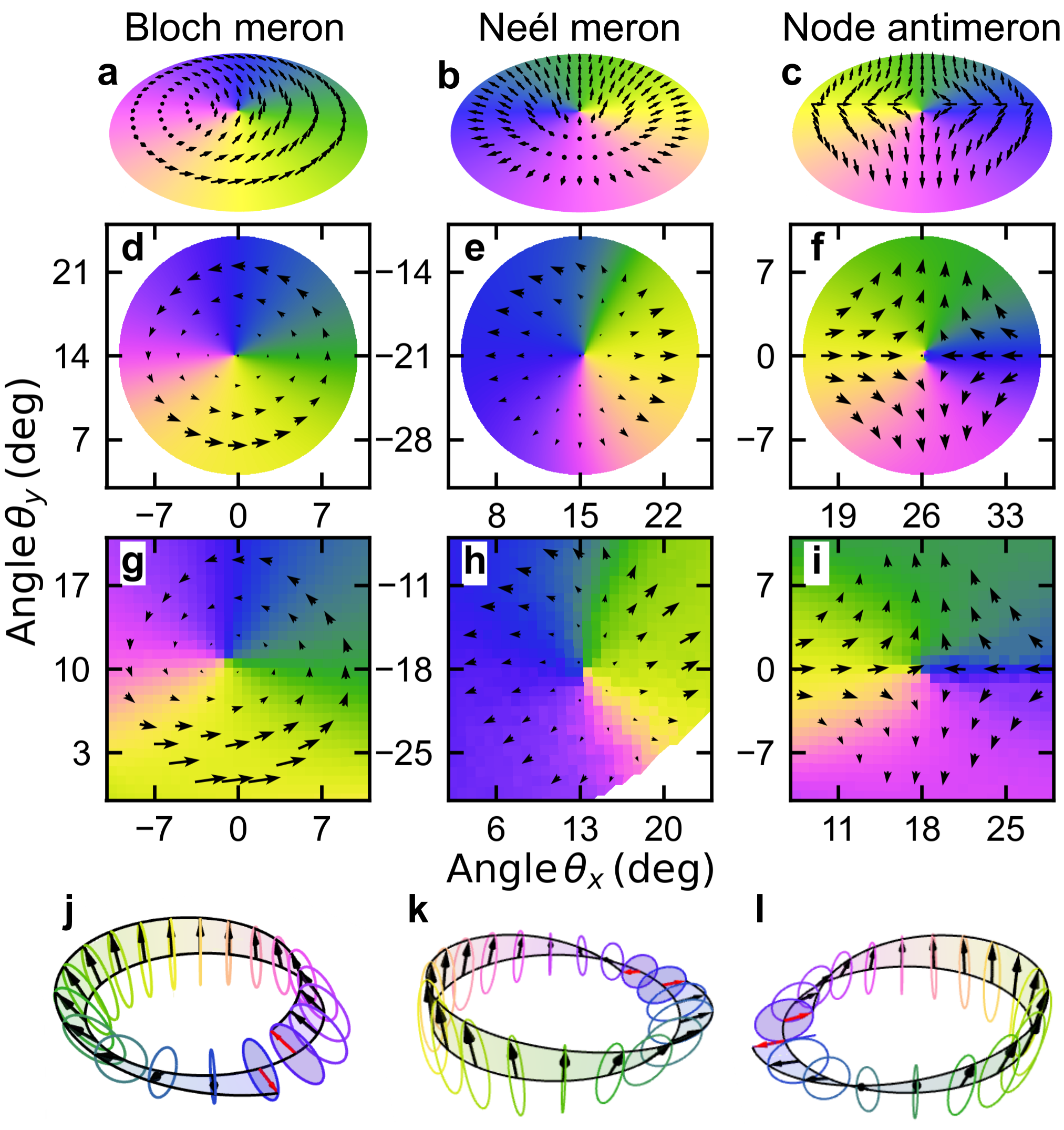}
    \caption{\textbf{Distribution of Stokes parameters, phase and polarization ellipse for three different type of merons.} \textbf{a-c} Distribution of vector of Stokes polarization parameters $\bm{s}$ and Stokes phase for Bloch meron,  Ne\'el meron and Node anti-meron  in ideal case, respectively, (see Section III in SM~\cite{Supplementary_Materials}).  \textbf{d-f} The same distribuiton obtained from Berreman and Schubert method and \textbf{g-i} for experimental data. \textbf{j}-\textbf{l}  Möbius strips created by the polarization ellipse along a contour surrounding the polarization singularity. The arrows indicate the major semi-axes of the ellipse, with the red arrow denoting the discontinuity in the major axis. The polarization ellipses are rotated by an angle of  $\varphi_{1,2}/2 + \pi/2$ relative to the axis tangent to the circle.}
    \label{Figure main - Three type of merons}
\end{figure}

Interestingly, the lemon-type C-points between the upper and lower branches possess different helicities from skyrmionic considerations~\cite{Shen_NatPhotRev2024}. Upper branch C-points show characteristics similar to first-order Bloch merons [see Fig.~\ref{Figure main - Three type of merons}\textbf{a},\textbf{d},\textbf{g}], whereas the lower branch exhibits first-order N\'eel merons [see Fig.~\ref{Figure main - Three type of merons}\textbf{b},\textbf{e},\textbf{h}]. The difference between these Bloch- and N\'eel-meron textures can also be appreciated from the two-dimensional Stokes divergence $\gamma_{1,2} = \partial_{k_x} s_{1} + \partial_{k_y} s_{2}$, where $\gamma_{1,2}\approx 0$ and $\neq 0$ close to the Bloch- and N\'eel-meron, respectively (see Figs.~S3 and S4 in the SM~\cite{Supplementary_Materials}). Additionally, we can change the C-points from Bloch to N\'eel type by simply adjusting the detuning $\Delta$ which flips the meron helicity from $0$ to $\pi/2$ (see Figs.~S3 and S4 in the SM~\cite{Supplementary_Materials}). Similar transformations between magnetic N\'eel skyrmion and Bloch skyrmionic bubbles have been observed in van der Waals ferromagnet Fe$_{3-\delta}$GeTe$_2$ \cite{Liu_(2023)_Adv.Sci.}. In those materials the meron type is determined by a sample thickness and iron concentration. In contrast, our momentum space optical cavity ``meron'' transformation is controlled in-situ by tuning the external voltage. Additionally, the electric field (represented by the polarization ellipses) around an odd number of C-points forms a Möbius strip \cite{Bliokh_(2021)_Phys.Fluids}, as illustrated in Figs.~\ref{Figure main - Three type of merons}\textbf{j}-\textbf{l} for the three types of merons present in the system. By implementing in-situ changes to the type of meron (from Bloch to Néel and vice versa), we can modify the orientation of the Möbius strip in reciprocal space and, for a given polarization, determine the wavevector at which the discontinuity of the major semi-axis of the polarization ellipse occurs (see the red arrows in Figs.~\ref{Figure main - Three type of merons}\textbf{j}-\textbf{l}).

At the DPs the half-charge polarization singularities resemble anti-merons but with undefined polarization at the center, hence we refer to it as a node anti-meron [see Fig.~\ref{Figure main - Three type of merons}\textbf{c},\textbf{f},\textbf{i}]. These textures have opposite vorticity compared to the Bloch and Néel merons and cannot be smoothly changed since they come from band degeneracy.

We performed the same experiment and numerical modeling for negative detuning ($\Delta < 0$), with the results presented in Fig.~S5. As predicted, under negative detuning, we observed only two C-points in the lower branch, with a polarization distribution similar to a Bloch meron (see Fig.~S6).

\begin{figure*}
\includegraphics{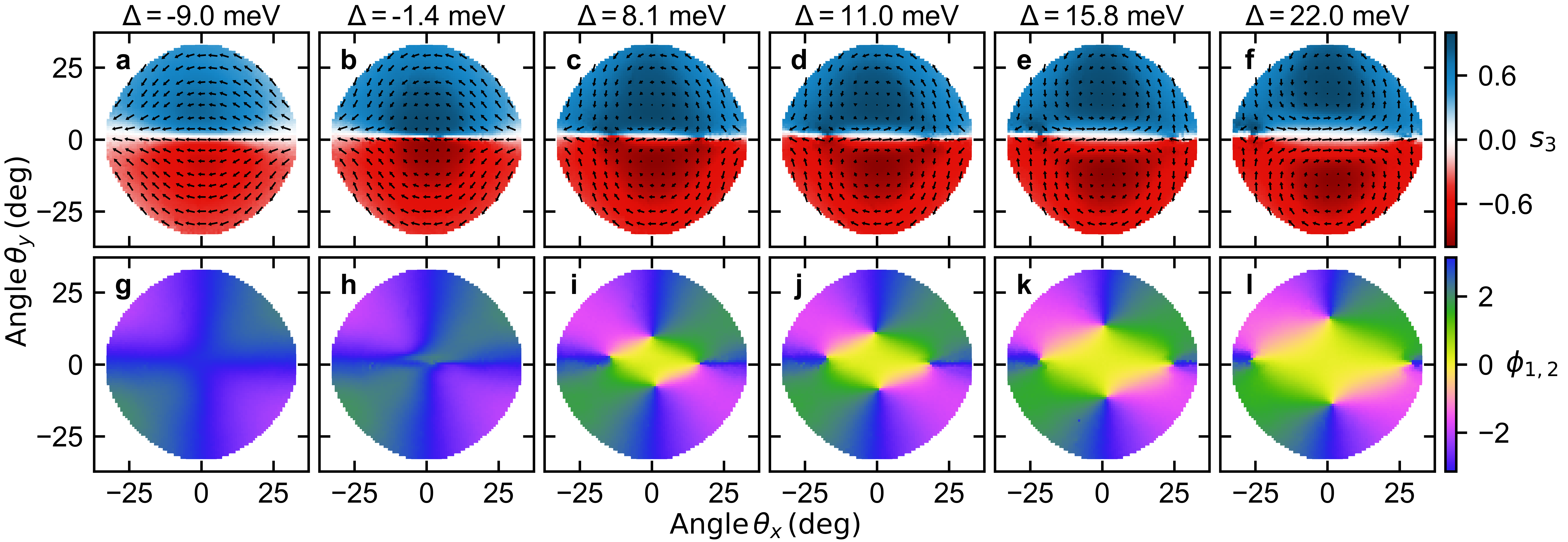}
  \caption{\textbf{Tunability of the polarization singularities} \textbf{a-f} Upper branch $s_3$ Stokes parameter for as a function of detuning with overlaid $\bm{s}_{\parallel} = \left[s_{1}, s_{2}\right]^{T}$ as black arrows. \textbf{g-i} Corresponding Stokes phase $\phi_{1,2} = \arg\left(s_{1} + is_{2}\right)$ showing changing position of the singularities in reciprocal space.}
  \label{Figure main - Tunability of C-points}
\end{figure*}

So far, we have demonstrated that in LCMC, we can alter both the number of C-points and their helicity (Bloch or N\'eel meron-like) by adjusting the detuning. Here, we further illustrate our ability to continuously tune the position of C-points in momentum space. Figure~\ref{Figure main - Tunability of C-points} depicts the $s_{3}$ Stokes polarization parameter, $\bm{s}_{\parallel}$ vector, and the Stokes phase $\phi_{1,2}$ for the upper branch for various detuning values in Sample A. The distribution of these parameters for the lower branch is presented in Fig.~S7 in \cite{Supplementary_Materials}. For negative detunings, no C-points are observed. However, with positive detuning values, two C-points appear, and importantly, their positions in reciprocal space can be smoothly adjusted by varying the detuning via external voltage control.

\section{Discussion}

We show that extreme birefringent liquid crystal microcavities can function as electrically tunable optical microscale devices for generation of multiple momentum-space polarization singularities including purely circularly-polarized points, C-points with quantized winding of the associated linear polarization vector. 

This offers a clear advantage over irreversibly nanopatterned photonic structures which must be carefully pre-designed to achieve the desired output or transmission properties with little room for in-situ adjustments. Given its electrical tunability, our device could prove useful to study in a controllable manner the evolution of geometric and dynamical phases~\cite{Bliokh_IOP2019}, for optical-based topological differential analysis~\cite{Zhu_NatComm2021}, non-diffracting space–time light bullets~\cite{Guo_LSA2021}, and potentially exploring the role of optical skyrmions in quantum entanglement~\cite{Ornelas_NatPhot2024}. 

Another exciting perspective of generating C-points in microcavities, instead of photonic crystal slabs, is the high Q-factor and small mode volumes of the microcavity which makes it convenient for lasing applications~\cite{Muszynski_PRAppl2022} and to access the strong light-matter coupling regime~\cite{Cilibrizzi_PRB2016}, even at room temperature~\cite{Lempicka_SciAdv2022}. The latter leads to enhanced nonlinear effects in out-of-equilibrium condensation of exciton-polaritons~\cite{Li_NatComm2022, Colas_LightSciAppl2015}. Furthermore, we have only studied the polarization structure of the transmitted cavity far field which is mostly paraxial. A new venture into liquid crystal cavity optics would be to study electrically tunable three-dimensional knotted C-lines in a nonparaxial field, similar to those generated using spatial light modulator technologies~\cite{Larocque_NatPhys2018}.  

In addition, this work shows that the birefringent microcavity is an excellent platform for investigating non-Hermitian photonic physics. The non-orthogonality of the cavity eigenmodes (non-Hermiticity) can be tuned by e.g. changing the number of pairs of layers in DBRs [see Section C in Methods] and strongly depends on the angle of the incident wave. Therefore, future studies should consider the impact of higher-order terms of the wave vector. Finally, this work shows that the description of the cavity is more complex than previously thought, making it a valuable platform for further developing Hamiltonians that describe non-Hermitian physics.

\section{Materials and methods}

\subsection{Information about sample}
Sample A consists of two DBRs composed of 6 pairs of layers $\mathrm{TiO_{2}}/\mathrm{SiO_{2}}$ with a central wavelength at 550~nm and $\mathrm{SiO}_{2}$ as a top layer. The DBRs were deposited on 25~nm ITO electrode layers on a glass substrate with the flatness of $\lambda/4$ at 633~nm. Antiparallel orienting layers (SE-130, Nissan Chem., Japan) were deposited on both substrates using the spin-coating method. The cavity was filled with a liquid crystal mixture $\mathrm{LC}2091^{\ast}$ in the nematic phase by capillary action. The sample consists dye (P580 -- Pyrromethene 580) in $\mathrm{CHCl_{3}}$ with mass concentration about 1\%.

Sample B consists of two DBRs composed of 6 pairs of $\mathrm{TiO_{2}}/\mathrm{SiO_{2}}$, with a central wavelength of 530~nm and $\mathrm{SiO}_{2}$ as a top layer. The ITO electrode, glass substrate, and orienting layers are the same as in sample A. Sample B is filled with the liquid crystal mixture $\mathrm{LC1892}^{\ast}$ in the nematic phase through capillary action.

\subsection{Data acquisition and analysis}
The experimental results were obtained by polarization-resolved tomography in transmission configuration. The broadband LED was used as a~light source. The automated setup of the quarter-wave and half-wave plates with a fixed linear polarizer was used to achieve desired polarization state of the incident light. The polarized light was focused on a sample using a~microscope objective with a numerical aperture of \mbox{$\mathrm{NA} = 0.7$}. The electrodes of the sample were connected to an external voltage source with a square waveform signal of 1 kHz frequency. Transmitted light was collected with a microscope objective with \mbox{$\mathrm{NA} = 0.75$} and focused on a slit of a monochromator equipped with a CCD camera with a set of reciprocal space imaging lenses. By automatized move of the last lens coupled with an acquisition of the CCD camera, 3-dimensional maps of energy and momentum in $k_x$ and $k_y$ directions were reconstructed for six polarizations (horizontal (H), vertical (V), diagonal (D), antidiagonal (A), right-hand circular ($\sigma^+$) or left-hand circular ($\sigma^-$)) and for different voltage values, corresponding to certain values of detuning.

In our data analysis, we fitted two Lorentzian functions $f_{\pm, i}\left(\lambda, \lambda_{\pm}, A_{\pm, i}\right) = A_{\pm, i}/(1+ ((\lambda - \lambda_{\pm})/\gamma_{\pm})^{2}$ to two orthogonally polarized spectra simultaneously, using the same width and peak position for both spectra while allowing the amplitudes to vary independently. This procedure yields two linewidths ($\gamma_{+}$ and $\gamma_{-}$) and energies ($\lambda_{+}$ and $\lambda_{-}$) (for the upper ($+$) and lower ($-$) branches) and four amplitudes ($A_{+,1}$ and $A_{+, 2}$ for the upper and $A_{-,1}$ and $A_{-, 2}$ lower branches in two orthogonal polarizations). The Stokes polarization parameters presented in Figs.~2 and 3 were calculated as $S_{\pm} = (A_{\pm, 1} - A_{\pm, 2})/(A_{\pm, 1} + A_{\pm, 2})$.

\subsection{Non-Hermitian modification of 2-mode Hamiltonian due to DBR modes}

One issue requiring clarification is the reason behind the discrepancy between the experimental findings the simplified $2\times2$ Hamiltonian approach. At both positive and negative detunings, the experiment and Berreman-and-Schubert method reveal two additional C-points in the lower branch that are not present in the $2\times2$ Hamiltonian solutions. Here, we argue that this discrepancy arises due to the interaction of the cavity modes with the Bragg modes of the DBRs. Figures~S8 and S9 in the SM~\cite{Supplementary_Materials} show the Stokes polarization parameters for the two branches calculated using the Berreman method for the photonic structure based on Sample A, but with 4 and 8 pairs of layers in the DBRs, respectively. As shown in these figures, the polarization pattern for the lower branch strongly depends on the number of pairs of layers in DBRs. For the Hermitian $2\times2$ Hamiltonian matrix $\mathcal{H}$ written in the linear basis, the normalized eigenvectors $\bm{q}_{\pm}$ ($+$ and $-$ denotes the upper and lower branch, respectively) are orthogonal, so $\left\langle \bm{q}_{\pm} \left| \bm{q}_{\pm} \right\rangle\right. = \delta_{\pm, \pm}$, where $\delta_{ij}$ denotes the Kronecker delta. The normalized Stokes polarization parameters defined as $s_{i, \pm} = \bm{q}^{\dag}\bm{\sigma}_{i}\bm{q}$, where $\bm{\sigma}$ is the same as in Eq.~\eqref{equation: 2x2 hamiltonian - simple version}, are orthogonal on the Poincar\'e sphere, which means that $s_{i, +} = -s_{i, -}$. For a Hermitian system $\left\langle \bm{s}_{+} \left| \bm{s}_{-} \right\rangle\right. = -1$, where $\bm{s}_{\pm} = \left[s_{1, \pm}, s_{2, \pm}, s_{3, \pm}\right]^{T}$. The numerical results show that with the increasing number of layers in DBRs, we observe an increase in the non-orthogonality in the system, defined as $\chi = \left(\left\langle \bm{s}_{+} \left| \bm{s}_{-} \right\rangle\right. + 1\right)/2$, so for fully orthogonal states $\chi = 0$, while for a fully non-orthogonal system $\chi = 1$. The increasing number of layers in the DBRs narrows the stopband, bringing the Bragg modes closer to the cavity modes. Consequently, this proximity enhances their interaction, resulting in an increase in $\chi$ and a greater impact of non-Hermiticity in the system under consideration. Figures~S10 and S11 in the SM~\cite{Supplementary_Materials} illustrate the values of $\chi$ for 4, 6, and 8 pairs of layers for sample A and sample B, with the remaining parameters consistent with those in Figs.~\ref{Figure main - Comparison between theory and experiment - positive detuning} and S5, respectively. 

Our findings underscore the limitations of the $2\times2$ Hamiltonian in describing LCMCs and highlight several important considerations for accurately describing polarization patterns in liquid crystal microcavities. Firstly, higher-order expressions, including up to fourth order or more, may be crucial. The fourth order term in the non-Hermitian Hamiltonian is crucial for obtaining two and four C-points in the upper and lower branches, respectively. However, this fourth order term is sometimes insufficient to accurately reproduce the distribution of the imaginary part of the energy for the two branches, requiring higher-order terms for proper modelling. For example, in a recent work \cite{Robin_Hu_(2023)_Phys.Rev.B} authors include higher-order wave vector terms (up to fourth order) but only in the imaginary component (which accounts for system losses). Secondly, a non-Hermitian Hamiltonian matrix is necessary to describe phenomena like EPs \cite{Krol_NatComm2022, Su_SciAdv2021} and non-orthogonal states on the Poincaré sphere effectively. Thirdly, the interaction between cavity modes and Bragg modes within distributed Bragg reflectors (DBRs) can significantly impact results, as proved by this work. Therefore, the $2\times2$ Hamiltonian should account for this interaction, possibly through perturbation theory or similar methods. 

Additionally, our study highlights the necessity for a comprehensive approach to developing a suitable $2\times2$ matrix that incorporates all the aforementioned components, rather than proposing Hamiltonians without proper derivation, as is often the case. In a recent paper  \cite{Oliwa_(2024)_Phys.rev.res.}, we introduced a method for deriving a general formula for the $2\times2$ non-Hermitian Hamiltonian for a birefringent cavity. However, this approach proves inadequate in this context because the Hamiltonian derived does not consider higher-order terms in the wave vector or the influence of the DBR modes, as analysis was confined to cavities resembling Fabry-Pérot resonators. Future work will focus on developing a model that accounts for these interactions.

\section{Acknowledgement}
This work was supported by the National Science Center, Poland, under the project No. and 2023/51/B/ST3/03025 (J.S.), 2021/43/I/ST5/02632 (M.P.), 2019/35/B/ST3/04147 (M.M.) and 2022/45/P/ST3/00467 (H.S.) co-funded by the European Union Framework Programme for Research and Innovation Horizon 2020 under the Marie Skłodowska-Curie grant agreement No. 945339. and the European Union’s Horizon 2020 program, through a FET Open research and innovation action under the grant agreements No. 964770 (TopoLight) (P.O., P.Ka., B.P.).

\section{Contributions}
P.O., P.Ka. and H.S. wrote the manuscript under supervision of J.S. M.P., M.K., P.O, P.Ka. and M.M. performed experiment. J.S and B.P. developed methodology of the experiment. P.M. and R.M. prepared liquid crystal microcavities under W.P. supervision, P.Ku. synthesised the liquid crystal mixture. P.O. and W.B. developed the theory with input from H.S. P.O. performed simulations and visualised the data. J.S, B.P., H.S. and W.P.  provided  financial support for the project. 
\newline

\section{Data availability}
The data generated in this study are available in the Dane Badawcze UW Repozytorium: https://doi.org/10.58132/XCJJ6Z.

\section{Conflict of interest}
The authors declare no competing interests.

\section{Keywords}
Momentum space polarization singularities, liquid crystal microcavity, spin-orbit coupling of light, meron polarization texture, polarization optics, polarization vortex

\section{References}
\bibliography{bibliography}

\end{document}


\title{Electrically tunable momentum space polarization singularities in liquid crystal microcavities}

\author{Przemys\l{}aw\,Oliwa}
\affiliation{Institute of Experimental Physics, Faculty of Physics, University of Warsaw, ul.~Pasteura 5, PL-02-093 Warsaw, Poland}
\author{Piotr Kapu\'sci\'nski}
\affiliation{Institute of Experimental Physics, Faculty of Physics, University of Warsaw, ul.~Pasteura 5, PL-02-093 Warsaw, Poland}
\author{Maria Pop\l{}awska}
\affiliation{Institute of Experimental Physics, Faculty of Physics, University of Warsaw, ul.~Pasteura 5, PL-02-093 Warsaw, Poland}
\author{Marcin Muszy\'nski}
\affiliation{Institute of Experimental Physics, Faculty of Physics, University of Warsaw, ul.~Pasteura 5, PL-02-093 Warsaw, Poland}
\author{Mateusz Kr\'ol}
\affiliation{Institute of Experimental Physics, Faculty of Physics, University of Warsaw, ul.~Pasteura 5, PL-02-093 Warsaw, Poland}
\author{Przemys\l{}aw Morawiak}
\affiliation{Institue of Applied Physics, Military University of Technology, ul.~gen.~Kaliskiego 2, PL-00-908 Warsaw, Poland}
\author{Rafa\l{} Mazur}
\affiliation{Institue of Applied Physics, Military University of Technology, ul.~gen.~Kaliskiego 2, PL-00-908 Warsaw, Poland}
\author{Wiktor Piecek}
\affiliation{Institue of Applied Physics, Military University of Technology, ul.~gen.~Kaliskiego 2, PL-00-908 Warsaw, Poland}
\author{Przemys\l{}aw Kula}
\affiliation{Institue of Chemistry, Military University of Technology, ul.~gen.~Kaliskiego 2, PL-00-908 Warsaw, Poland}
\author{Witold Bardyszewski}
\affiliation{Institute of Theoretical Physics, Faculty of Physics, University of Warsaw, ul.~Pasteura 5, PL-02-093 Warsaw, Poland}
\author{Barbara Pi\k{e}tka}
\affiliation{Institute of Experimental Physics, Faculty of Physics, University of Warsaw, ul.~Pasteura 5, PL-02-093 Warsaw, Poland}
\author{Helgi Sigur{\dh}sson}
\email{Helgi.Sigurdsson@fuw.edu.pl}
\affiliation{Institute of Experimental Physics, Faculty of Physics, University of Warsaw, ul.~Pasteura 5, PL-02-093 Warsaw, Poland}
\affiliation{Science Institute, University of Iceland, Dunhagi 3, IS-107, Reykjavik, Iceland}
\author{Jacek Szczytko}
\email{Jacek.Szczytko@fuw.edu.pl}
\affiliation{Institute of Experimental Physics, Faculty of Physics, University of Warsaw, ul.~Pasteura 5, PL-02-093 Warsaw, Poland}
\maketitle
\begin{figure}
    \centering
    \includegraphics{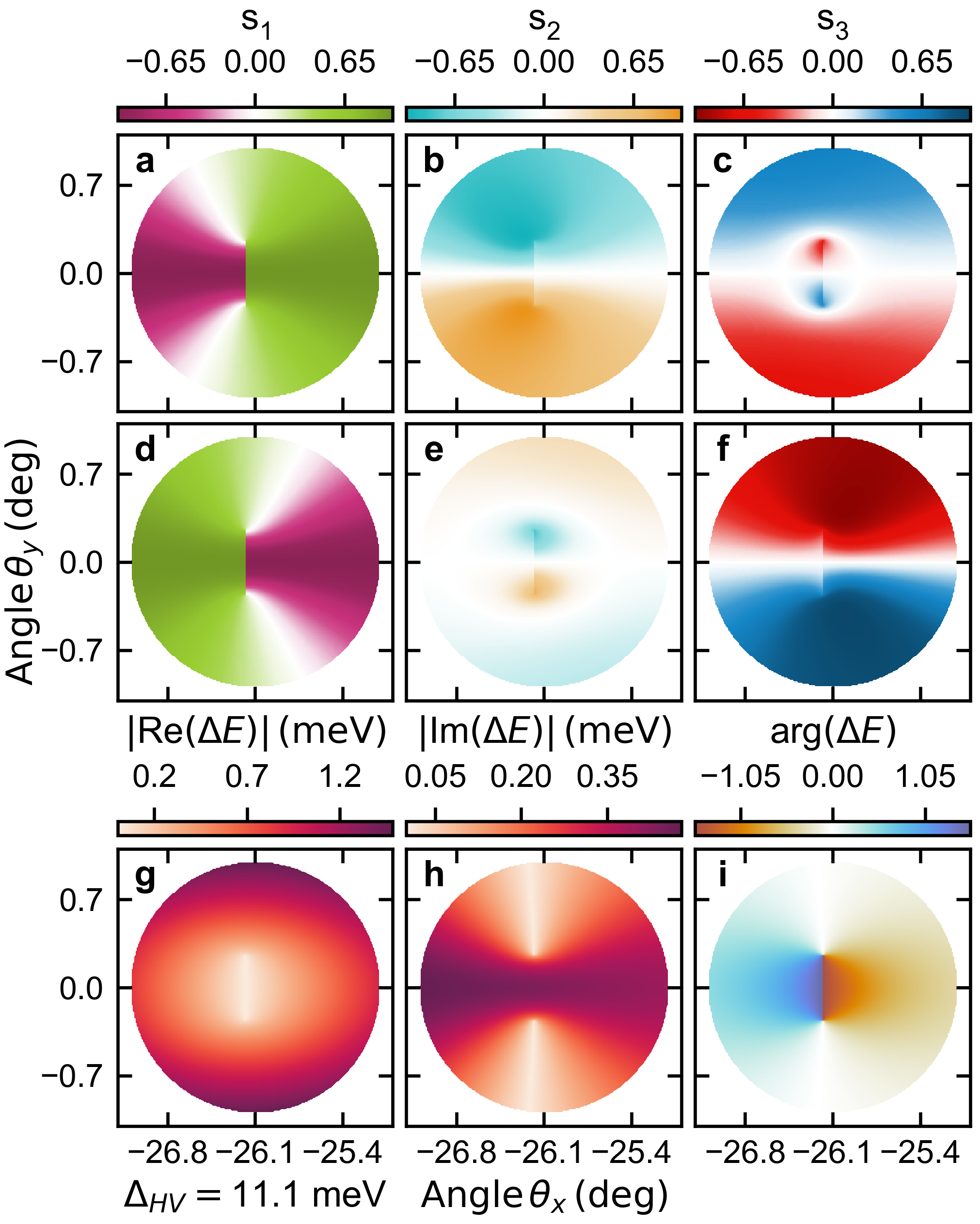}
    \caption{Polarization patterns and difference between energies for the two cavity photon branches close to the EPs, corresponding to our analysis in Section II.B in the main manuscript. \textbf{a}-\textbf{c} and \textbf{d}-\textbf{f} present polarization patterns for upper and lower branches, respectively. \textbf{g} and \textbf{h} presents the absolute value of the real and imaginary part difference between energies for two branches, respectively. \textbf{i} presents the argument of complex number -- difference between energies for both branches. The $\Delta_{HV}$ denote the difference between energies for two modes for perpendicular incident wave.}
    \label{SI - figure: Polarization patterns close to the left exceptional points}
\end{figure} 

\begin{figure}
    \centering
    \includegraphics{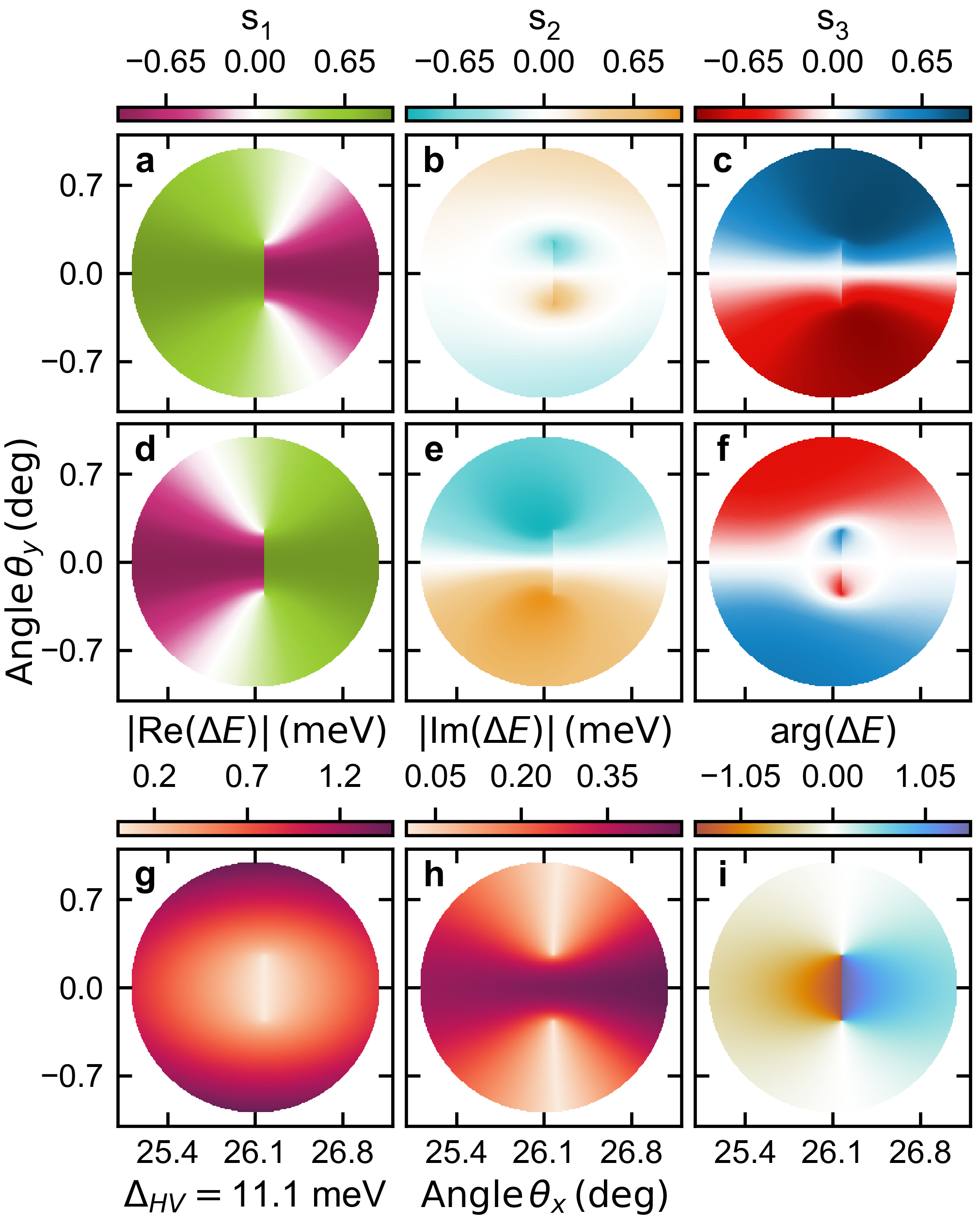}
    \caption{Polarization patterns and difference between energies for the two cavity photon branches close to the EPs, corresponding to our analysis in Section II.B in the main manuscript. \textbf{a}-\textbf{c} and \textbf{d}-\textbf{f} present polarization patterns for upper and lower branches, respectively. \textbf{g} and \textbf{h} presents the absolute value of the real and imaginary part difference between energies for two branches, respectively. \textbf{i} presents the argument of complex number -- difference between energies for both branches. The $\Delta_{HV}$ denote the difference between energies for two modes for perpendicular incident wave.}
    \label{SI - figure: Polarization patterns close to the rigth exceptional points}
\end{figure}

\section{2-mode Hamiltonian}
\subsection{Parameters of 2-mode Hamiltonian}
In each equation, we used the following notation. The subscripts $s = X$ and $s = Y$ denote the horizontally and vertically polarized cavity mode, respectively. The subscripts $X\left(m\right)$ and $Y\left(m\right)$ denote the $m$-th horizontally and vertically polarized mode, respectively.   
The value of parameters of $2\times2$ Hamiltonian presented in Eq.~(2) are equal to:
\begin{subequations}
    \label{equation: N+1,N regime parameters}
    \begin{equation}
        \omega_0 = \mathrm{Re}\left(\frac{\overset{o}{\omega}_{X\left(m + 1 \right)} + \overset{o}{\omega}_{Y\left(m\right)}}{2} \right)
    \end{equation}
    \begin{equation}
        \Gamma_0 = \mathrm{Im} \left(\frac{\overset{o}{\omega}_{X\left(m+1\right)} + \overset{o}{\omega}_{Y\left(m\right)}}{2} \right)
    \end{equation}
    \begin{equation}
        \Delta = \mathrm{Re} \left(\frac{\overset{o}{\omega}_{X \left(m + 1\right)} - \overset{o}{\omega}_{Y \left(m\right)}}{2}\right)
    \end{equation}
    \begin{equation}
        \delta \Gamma = \mathrm{Im} \left(\frac{\overset{o}{\omega}_{X \left(m + 1 \right)} - \overset{o}{\omega}_{Y \left( m \right)}}{2} \right)
    \end{equation}
\end{subequations}
The $\overset{o}{\omega}_{sm}$ denote the resonant frequency for $s$ polarized mode with $m$-th mode number, equal to~\cite{Oliwa_(2024)_Phys.rev.res.}:
\begin{equation} \label{eq.om0}
    \overset{o}{\omega}_{sm} = \frac{c m_{s} \pi}{n_{s} L} \left( 1 - \frac{1}{L \zeta_{s}} + \frac{1}{L^{2} \zeta_{s}^{2}} - \frac{i n_{a} m_{s} \pi}{2L^{2} \zeta_{s}^{2} n_{s}} \right) 
\end{equation}
where $\zeta_{s}$ denote the "strength of $\delta$-mirrors" for $s$ polarization, $c$ is the speed of light, $L$ is the cavity thickness, $n_s$ is the cavity refractive index, and $n_a=1$ is the outside refractive index. The terms quadratic in transverse wave vector in Eq.~(3) in the main text have the following form:
\begin{subequations}
     \begin{equation}
        \frac{1}{m_{x}} = \frac{1}{2m_{0}} \left( \frac{a_{x}}{\eta} + b_{x}\eta \right)
    \end{equation}
    \begin{equation}
        \frac{1}{m_{y}} = \frac{1}{2m_{0}} \left(\frac{a_{y}}{\eta} + b_{y}\eta\right)
    \end{equation}
    \begin{equation}
        \delta_{x} = \frac{\hbar}{4m_{0}} \left(\frac{a_{x}}{\eta} - b_{x}\eta\right)
    \end{equation}
    \begin{equation}
        \delta_{y} = \frac{\hbar}{4m_{0}} \left(\frac{a_{y}}{\eta} - b_{y}\eta\right)
    \end{equation}
\end{subequations}
The masses $m_{x}$ and $m_{y}$ describe the mean value of the curvature of two parabolas for two orthogonally polarized modes. The $\delta_{x}$ and $\delta_{y}$ describe the difference between these curvatures, which occur due to the presence of \mbox{$TE-TM$} splitting, which is caused by strong birefringent of the medium inside the cavity. The other parameters are explained below. In the considered case, we also have one term linear in wave vector, whose coefficient is equal to:
\begin{equation}
    \label{eq56e}
    \alpha = -\frac{\hbar}{4m_0}\frac{\varepsilon_{xz}}{n_{X} n_{Y} \varepsilon_{zz}}Q_{X \left(m + 1\right), Y \left( m \right)}.
\end{equation}
This is the so-called Rashba-Dresselhaus coefficient~\cite{Rechcinska_Science2019}. The refractive indices for each polarization are \mbox{$n_{s} = \sqrt{\tilde{\varepsilon}_{ss}}$}, where $\tilde{\varepsilon}_{ij} = \varepsilon_{ij} - {\varepsilon_{iz}\varepsilon_{zj}} /{\varepsilon_{zz}}$ and $\varepsilon_{ij}$ denotes the element of dielectric tensor. The $Q_{sm,s^{\prime}, m^{\prime}}$ describe the overlap between a mode and the derivative of another mode along the optical axis of the cavity:
\begin{multline}
    Q_{sm,s^{\prime}m^{\prime}} = \frac{n_sn_{s^{\prime}}}{c^2}\int_0^L \left(\bm E_{sm}\left(z \right) \right)^T\partial_z \bm {E}_{s^{\prime}m^{\prime}} \left(z \right)\, \mathrm{d}z \approx\\ 
    \approx 
    \begin{cases}
        0 & \text{if $m + m^{\prime}$ even,} \\
        \frac{4mm^{\prime}}{L (m^2-\left(m^{\prime}\right)^2)} + \frac{8(m^3\Delta_{s^{\prime}m^{\prime}} - \left(m^{\prime}\right)^3\Delta_{sm})}{\pi (m^2-\left(m^{\prime}\right)^2)^2} &  \text{if $m+m^{\prime}$ odd} \\
    \end{cases}
\end{multline}
The rest mass of photon in a cavity $m_{0}$ and the normalization $\eta$ are equal to:
\begin{subequations}
    \begin{equation}
        m_{0} = \frac{\hbar \sqrt{\overset{o}{\omega}_{X \left( m + 1 \right)} \overset{o}{\omega}_{Y \left( m \right)}}}{c^{2}}
    \end{equation}
    \begin{equation}
        \eta = \sqrt{\frac{\overset{o}{\omega}_{X \left( m + 1\right)}}{\overset{o}{\omega}_{Y \left( m \right)}}}
    \end{equation}
\end{subequations}
The $a_{x}$, $a_{y}$, $b_{x}$ and $b_{y}$ parameters are defined by the elements of $\bm{\varepsilon}$ matrix, which describes the material inside the cavity:
\begin{subequations}
    \begin{equation}
        a_{x} = \frac{1}{\tilde{\varepsilon}_{xx}} \left( \frac{\varepsilon_{xx}}{\varepsilon_{zz}} + 4\frac{\varepsilon_{xz}^{2}}{\varepsilon_{zz}}f_{X\left( m + 1 \right), X \left( m + 1\right)} \left( X \right) \right)
    \end{equation}
    \begin{equation}
        b_{x} = \frac{1}{\varepsilon_{yy}}
    \end{equation}
    \begin{equation}
        a_{y} = \frac{1}{\tilde{\varepsilon}_{xx}} \left( 1 + \frac{\varepsilon_{xz}^{2}}{\varepsilon_{zz}^{2}} \left( 1 + f_{X \left(m + 1\right), X \left( m+1 \right)} \left( Y \right) \right) \right)
    \end{equation}
    \begin{equation}
        b_{y} = \frac{1}{\varepsilon_{yy}} \left( \frac{\varepsilon_{yy}}{\varepsilon_{zz}} + \frac{\varepsilon_{xz}^{2}}{\varepsilon_{zz}^{2}}f_{Y \left(m\right), Y \left( m \right) }(X) \right)
    \end{equation}
\end{subequations}
where the $f_{sm, s^{\prime} m^{\prime}} (s^{\prime\prime})$ describes the impact from the other states present in the system, which is taken into account by perturbation theory. The first order correction are the terms quadratic in the wave vector and it is equal to:
\begin{multline}
    f_{sm, s^{\prime} m^{\prime}}\left(s^{\prime \prime} \right) = -{\sum_{m^{\prime \prime}}}^{\prime} \frac{c^2Q_{sm,s''m''}Q_{s''m'',s'm'}} {4\overset{o}{\omega}_{s''m''}n_{s''}^2}   \\  
    \times \left(\frac{1} {\overset{o}{\omega}_{sm} - \overset{o}{\omega}_{s''m''}} + \frac{1} {\overset{o}{\omega}_{s'm'} - \overset{o}{\omega}_{s''m''}}\right)
\end{multline}
The sum ${\sum_{m^{\prime\prime}}}^{\prime}$ is over all states which are different than the set of states belonging to 2-mode Hamiltonian, which in this case are equal to $X\left(m + 1\right)$ and $Y\left(m\right)$. Due to taking into account the impact from other states, the basis of the non-Hermitian Hamiltonian in Eq.~(2) from the main article is different than standard one in which one mode is horizontally polarized and the second is vertically polarized. In this case, the relation between the electric field in rotated basis $\bm{E}_{sm}^{\prime} \left(z\right)$ and in standard linear basis $\bm{E}_{sm} \left(z\right)$ is following:
\begin{figure}[!t]
    \centering
    \includegraphics{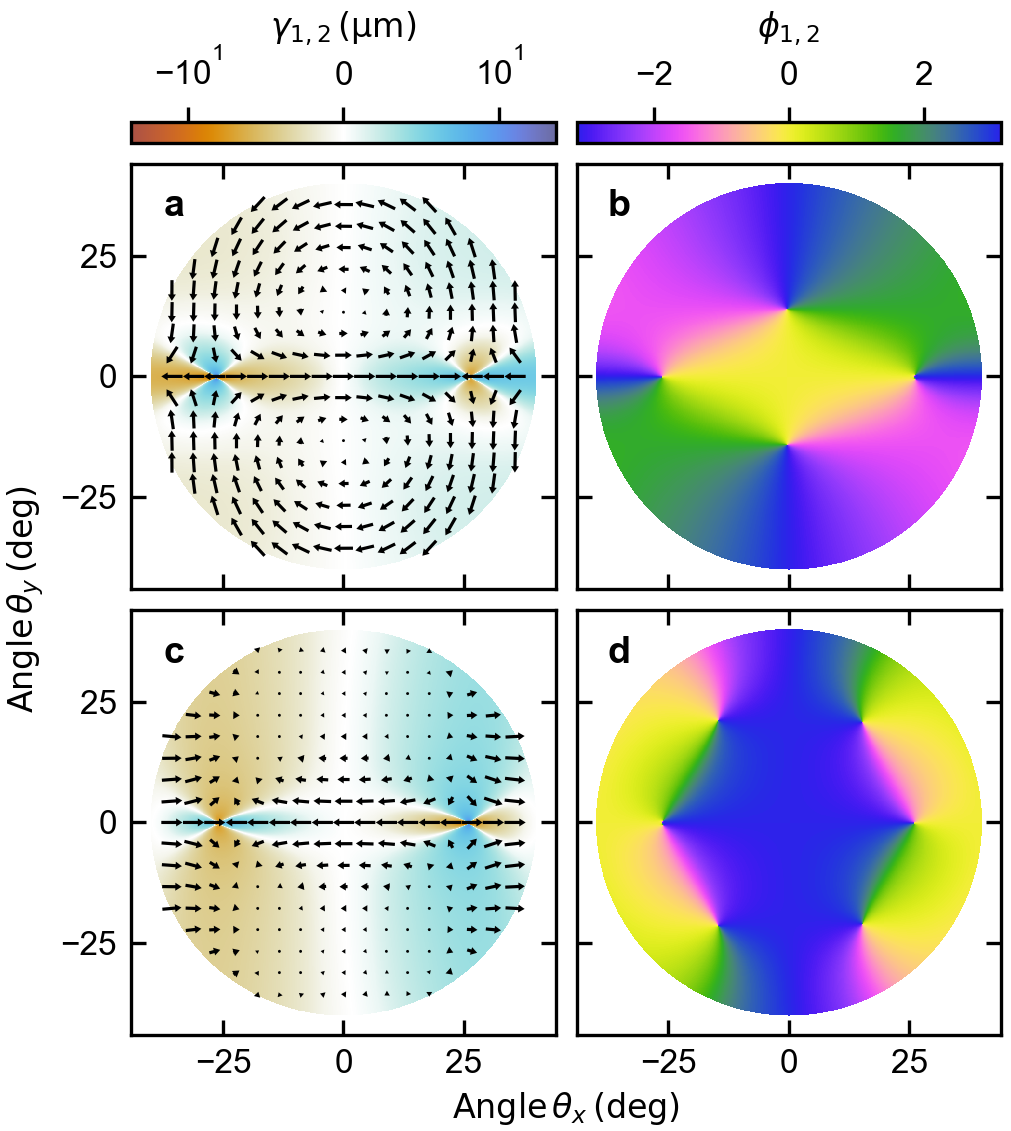}
    \caption{Two dimensional divergence and Stokes phase for $\Delta = 11.1\,\mathrm{meV}$ for Sample A ($\theta = 31.4^{\circ}$). \textbf{a} and \textbf{c}~two dimensional divergence $\gamma_{1,2}$ for upper \textbf{a} and lower \textbf{c}~branch, respectively. The arrows denote the $\bm{s}_{\parallel}$ vector. \textbf{b}~and \textbf{d} Stokes phase for upper and lower branch, respectively.}
    \label{SI - figure: Divergence and Stokes phase for Sample A high positive detuning}
\end{figure}
\begin{equation}
    \bm{E}_{sm}^{\prime} \left(z \right) = {\sum_{s^{\prime} m^{\prime}}}^{\prime} \left( \exp \left(\mathbf{R}\right)\right)_{s^{\prime}m^{\prime}} \bm{E}_{s^{\prime} m^{\prime}}
\end{equation}
where the summation is over all states different than the states belonging to the 2-mode Hamiltonian. The rotation operator $\mathbf{R}$ is defined as:
\begin{equation}
    \mathbf{R}_{sm, s^{\prime} m^{\prime}} =
    \begin{cases}
        -\frac{\mathcal{H}^{\left( 1 \right)}_{sm, s^{\prime} m^{\prime}}} {\overset{o}{\omega}_{sm} - \overset{o}{\omega}_{s^{\prime} m^{\prime}}} & \mathrm{if \,} \left(1\right) \\
        0 & \mathrm{if\,} \left(2\right) 
    \end{cases}
    \label{matrix R}
\end{equation}
where $\left(1\right)$ denotes the case in which $\bm{E}_{sm}$ and $\bm{E}_{s^{\prime} m^{\prime}}$ belong to different subsets of the basis set and $\left(2\right)$ denotes the opposite case. The $\mathcal{H}^{\left( 1 \right)}_{sm, s^{\prime} m^{\prime}}$ defines the linear term in $\bm{k}$ in multimode Hamiltonian in \cite{Oliwa_(2024)_Phys.rev.res.}, so in this case it is equal to:
\begin{equation}
    \mathcal{H}^{\left( 1 \right)}_{sm, s^{\prime} m^{\prime}} = \chi_{sm, s^{\prime} m^{\prime}} \mathbf{A}_{ss^{\prime}} Q_{sm, s^{\prime} m^{\prime}}
    \label{equation: linear term in Hamiltonian matrix}
\end{equation}
where $\chi_{sm, s^{\prime} m^{\prime}} = c^{2}/2n_{s}n_{s^{\prime}}\sqrt{\overset{o}{\omega}_{sm}\overset{o}{\omega}_{s^{\prime}m^{\prime}}}$ and $\mathbf{A}$ is equal to:
\begin{equation} \label{eq.Amatrix}
    \mathbf{A} = -i\frac{\varepsilon_{xz}}{\varepsilon_{zz}}
    \begin{bmatrix}
        2k_{x} & k_{y} \\ 
        k_{y} & 0 
    \end{bmatrix}
\end{equation}
\begin{figure}[!t]
    \centering
    \includegraphics{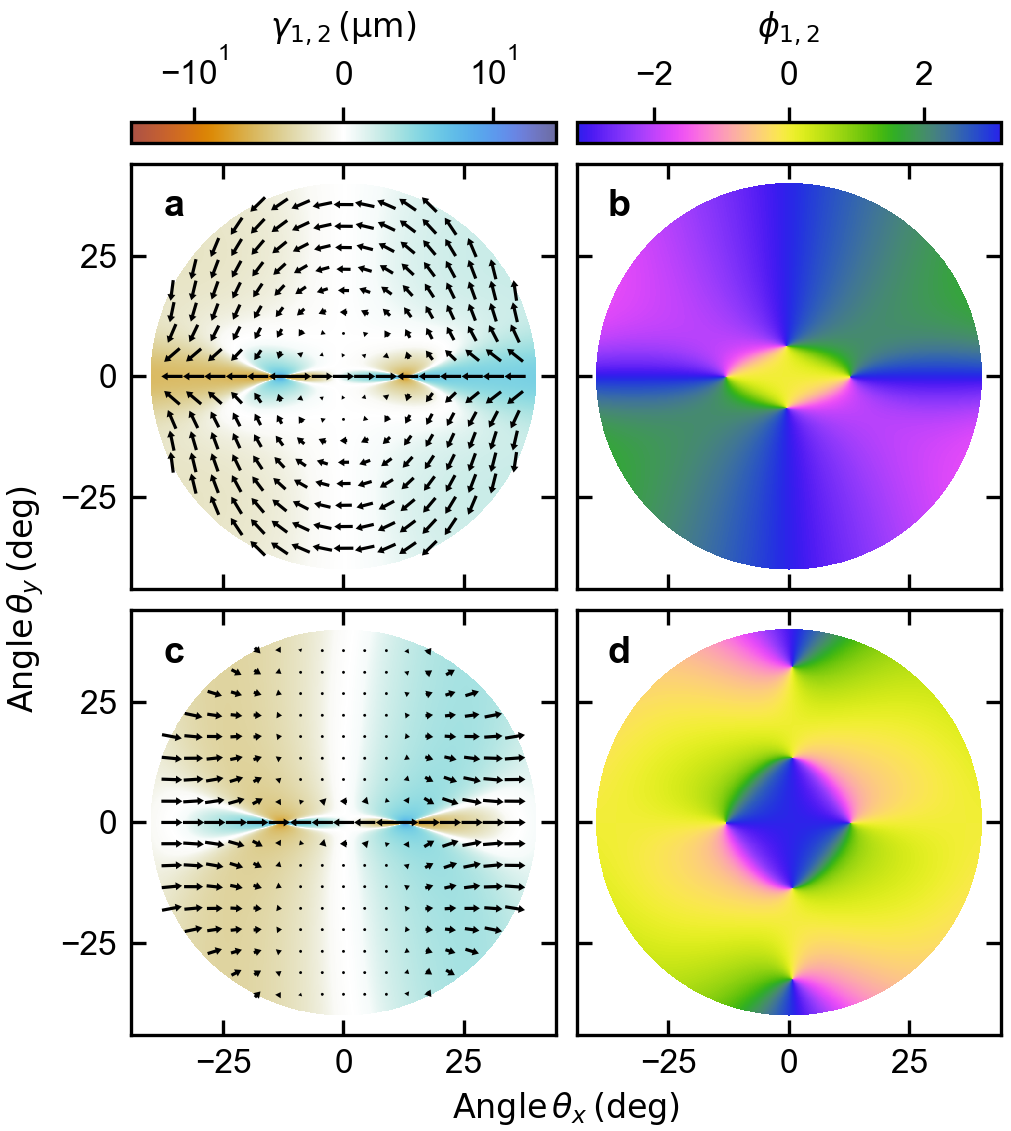}
    \caption{Two dimensional divergence and Stokes phase for $\Delta = 2.4\,\mathrm{meV}$ for Sample A ($\theta = 29.5^{\circ}$). \textbf{a} and \textbf{c}~two dimensional divergence $\gamma_{1,2}$ for upper \textbf{a} and lower \textbf{c}~branch, respectively. The arrows denote the $\bm{s}_{\parallel}$ vector. \textbf{b}~and \textbf{d} Stokes phase for upper and lower branch, respectively.}
    \label{SI - figure: Divergence and Stokes phase for Sample A low positive detuning}
\end{figure}
\subsection{Derivation of position of C-points}
The Hamiltonian presented in the main text in Eq.~(2) has the following two eigenvectors:
\begin{equation} \label{eq.q_orig}
    \bm{q}_{\pm} = 
    \begin{bmatrix}
        h_{1} \pm \sqrt{h_{1}^{2} + h_{3}^{2}} \\ 
        i h_{3}
    \end{bmatrix}
\end{equation}
where $h_{0}$, $h_{1}$ and $h_{3}$, as shown in Eq.~(3), are equal to:
\begin{figure*}[]
    \centering
    \includegraphics{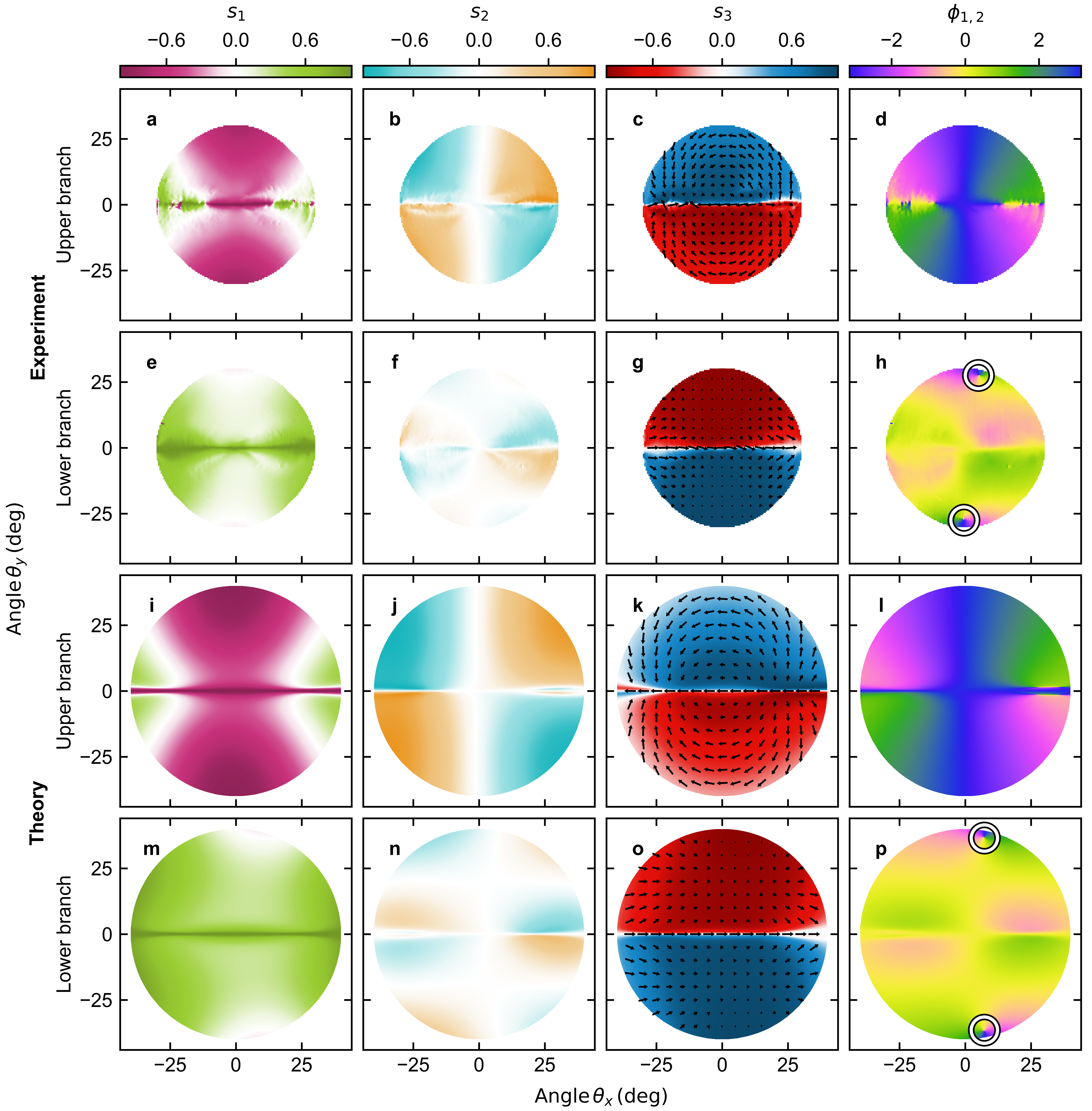}
    \caption{Stokes parameters and phase in negative detuning. \textbf{a-h} Experimentally and \textbf{i-p} theoretically obtained Stokes parameters \textbf{a}, \textbf{e}, \textbf{i}, \textbf{m} $s_1$, \textbf{b}, \textbf{f}, \textbf{j}, \textbf{n}, $s_2$, \textbf{c}, \textbf{g}, \textbf{k}, \textbf{o} $s_3$, and \textbf{d}, \textbf{h}, \textbf{l}, \textbf{p} Stokes phase $\phi_{1,2} = \mathrm{arg}\left(s_1 + i s_2\right)$ for \textbf{a-d, i-l} upper and \textbf{e-h, m-p} lower branch in the case of the positive detuning of $2\hbar \Delta = \hbar \mathrm{Re}\left( \omega_{H \left(m + 1\right)} - \omega_{V \left(m\right)} \right) = -1.7 \,\mathrm{meV}$. The black arrows on the $s_3$ map correspond to $\bm{s}_{\parallel} = \left[s_{1}, s_{2} \right]^{T}$. The white circles on the Stokes phase $\phi_{1,2}$ maps mark the position of the C-points.}
    \label{Figure main - Comparison between experiment and theory - negative detuning}
\end{figure*}
\begin{subequations}
\label{SI equation: all parameters of Hamiltonian}
    \begin{equation}
        h_{0} = \omega_{0} + i \Gamma_{0} + \frac{\hbar k_{x}^{2}}{2m_{x}} + \frac{\hbar k_{y}^{2}}{2m_{y}},
    \end{equation}
    \begin{equation}
        h_{1} = \Delta + i \delta \Gamma + \delta_x k_{x}^{2} + \delta_{y} k_{y}^{2},
        \label{SI equation: h1 parameter of Hamiltonian}
    \end{equation}
    \begin{equation}
        h_{3} = -2\alpha k_{y}.
    \end{equation}
\end{subequations}
In our analysis, $\mathrm{Im} \left(h_{1} \right) \neq 0$ and $\mathrm{Im} \left( h_{3} \right) \neq 0$, which results in a complicated expression for $\bm{q}^{\dag}_\pm$. However, when deriving the positions of the C-points, we perform the calculations in the Hermitian limit ($\zeta_{i} \rightarrow \pm \infty$), where $\mathrm{Im} \left(h_{1} \right) = 0$ and $\mathrm{Im} \left( h_{3} \right) = 0$, and $\bm{q}^{\dag}$ is simply given by:
\begin{equation}
    \bm{q}_{\pm}^{\dag} = 
    \begin{bmatrix}
        h_{1} \pm \sqrt{h_{1}^{2} + h_{3}^{2}}, - i h_{3}
    \end{bmatrix},
    \label{SI - equation: eigenvectors}
\end{equation}
which is just the conjugate transpose of Eq.~\eqref{eq.q_orig}. In the Hermitian limit, as $\Delta_{sm} \rightarrow 0$, the formulas for $\omega_{sm}$, $\chi_{sm, s^{\prime} m^{\prime}}$, and $Q_{sm, s^{\prime} m^{\prime}}$ simplify to the followings:
\begin{figure}[!t]
    \centering
    \includegraphics{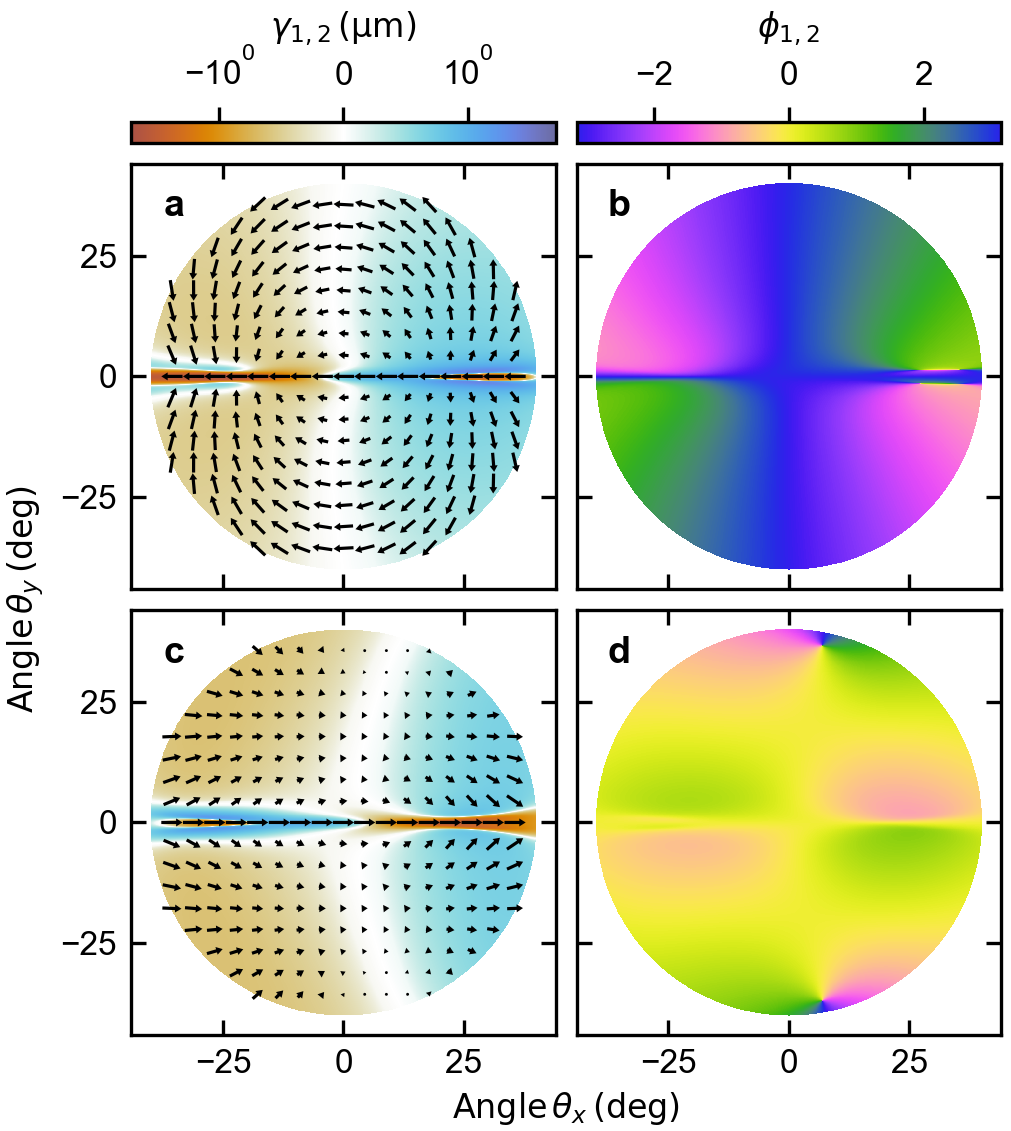}
    \caption{Two dimensional divergence and Stokes phase for $\Delta = -1.7\,\mathrm{meV}$ for Sample B ($\theta = 33.5^{\circ}$). \textbf{a} and \textbf{c}~two dimensional divergence $\gamma_{1,2}$ for upper \textbf{a} and lower \textbf{c}~branch, respectively. The arrows denote the $\bm{s}_{\parallel}$ vector. \textbf{b}~and \textbf{d} Stokes phase for upper and lower branch, respectively.}
    \label{SI - figure: Divergence and Stokes phase for Sample B negative detuning}
\end{figure}
\begin{subequations}
    \begin{equation}
        \omega_{sm} = \frac{c m \pi}{L n_{s}}
    \end{equation}
    \begin{equation}
        \chi_{sm, s^{\prime} m^{\prime}} = \frac{cL}{2\pi \sqrt{m m^{\prime} n_{s} n_{s^{\prime}}}}
    \end{equation}
    \begin{equation}
        Q_{sm, s^{\prime} m^{\prime}} =
    \begin{cases}
        0 & \text{if $m+m^{\prime}$ even,} \\
        \frac{4mm^{\prime}}{L (m^2-\left(m^{\prime}\right)^2)} &  \text{if $m+m^{\prime}$ odd} \\
    \end{cases}
    \end{equation}
\end{subequations}
thus, the term $\mathcal{H}_{sm, s^{\prime} m^{\prime}}^{(1)}$, defined in Eq.~\eqref{equation: linear term in Hamiltonian matrix}, can be simplified to the following formula:
\begin{equation}
    \mathcal{H}_{sm, s^{\prime} m^{\prime}}^{(1)} = \frac{2 c \tilde{\delta}_{m m^{\prime}}}{\pi \left(m^{2} - \left(m^{\prime}\right)^{2} \right)} \sqrt{\frac{m m^{\prime}} {n_{s} n_{s^{\prime}}}} \mathbf{A}_{s s^{\prime}},
\end{equation}
where $\tilde{\delta}_{m m^{\prime}}$ equals 1 if $m + m^{\prime}$ is an odd number, and 0 otherwise. Using these simplifications, we obtain a new formula for the nonzero elements of the rotation matrix $\mathbf{R}_{sm, s^{\prime} m^{\prime}}$:
\begin{equation}
    \mathbf{R}_{sm, s^{\prime} m^{\prime}} = -\frac{2L\tilde{\delta}_{m m^{\prime}} \sqrt{m m^{\prime} n_{s} n_{s^{\prime}}}}{\pi^{2} \left(m^{2} - \left(m^{\prime}\right)^{2}\right) \left(m n_{s^{\prime}} - m^{\prime} n_{s} \right)} \mathbf{A}_{s s^{\prime}}
    \label{equation: rotation matrix in hermitian limit}
\end{equation}
This formula can be used only if the $\bm{E}_{sm}$ and $\bm{E}_{s^{\prime}m^{\prime}}$ states belong to different subsets of the basis, as explained in Eq.~\eqref{matrix R}. In this case, the formula for the unnormalized Stokes polarization parameter takes the following form:
\begin{multline}
    S_{i, \pm} = \sum_{s, s^{\prime}} q_{\pm,s}^{\ast} q_{\pm, s^{\prime}} \left(\bm{\sigma}_{i, s s^{\prime}} \vphantom{ \sum_{s^{\prime \prime} m^{\prime \prime}}}\right. \\ + \left.\sum_{s^{\prime \prime} m^{\prime \prime}} \mathbf{R}_{s^{\prime \prime} m^{\prime \prime}, s m}^{\ast} \bm{\sigma}_{i, s^{\prime \prime} s^{\prime}} + \mathbf{R}_{s^{\prime \prime} m^{\prime \prime}, s^{\prime} m^{\prime}} \bm{\sigma}_{i, s s^{\prime \prime}}\right)
    \label{SI - equation: polarization parameter general formula}
\end{multline}
where $\bm{\sigma} = \left(\bm{\sigma}_{z}, \bm{\sigma}_{x}, \bm{\sigma}_{y} \right)$ as the considered Hamiltonian in written in the linear basis. The $\mathbf{A}$ matrix is anti-Hermitian and symmetric, so $\mathbf{A}_{s s^{\prime}}^{\ast} = -\mathbf{A}_{s s^{\prime}} = -\mathbf{A}_{s^{\prime} s}$ and \mbox{$\mathbf{R}_{sm, s^{\prime} m^{\prime}}^{\ast} = -\mathbf{R}_{sm, s^{\prime} m^{\prime}}$}. To simplify the formula for the Stokes polarization parameter, we consider the rotated Pauli matrix defined as:
\begin{multline}
        \tilde{\bm{\sigma}}_{i, s s^{\prime}} = \bm{\sigma}_{i, s s^{\prime}} + \\ 
        +\sum_{s^{\prime \prime} m^{\prime \prime}} \mathbf{R}_{s^{\prime \prime} m^{\prime \prime}, s^{\prime} m^{\prime}} \bm{\sigma}_{i, s s^{\prime \prime}} - \mathbf{R}_{s^{\prime \prime} m^{\prime \prime}, s m} \bm{\sigma}_{i, s^{\prime \prime} s^{\prime}} 
    \label{SI - equation: general pauli matrix}
\end{multline}
Substituting the formula from Eq.~\eqref{equation: rotation matrix in hermitian limit} into Eq.~\eqref{SI - equation: general pauli matrix} gives:
\begin{multline}
    \tilde{\bm{\sigma}}_{i, ss^{\prime}} = \bm{\sigma}_{i, ss^{\prime}} + \\ +
    \frac{2L}{\pi^{2}}\sum_{s^{\prime\prime} m^{\prime\prime}} \frac{\sqrt{m^{\prime\prime} m n_{s^{\prime\prime}} n_{s}} \mathbf{A}_{s^{\prime\prime}s}\tilde{\delta}_{m^{\prime\prime} m}}{\left(\left(m^{\prime\prime}\right)^{2} - m^{2}\right) \left(m^{\prime\prime} n_{s} - m n_{s^{\prime\prime}}\right)}\bm{\sigma}_{i, s^{\prime\prime}s^{\prime}} - \\ +
    \frac{\sqrt{m^{\prime\prime} m^{\prime} n_{s^{\prime\prime}} n_{s^{\prime}}} \mathbf{A}_{s^{\prime\prime}s^{\prime}} \tilde{\delta}_{m^{\prime\prime} m^{\prime}}}{\left(\left(m^{\prime\prime}\right)^{2} - \left(m^{\prime}\right)^{2}\right) \left(m^{\prime\prime} n_{s^{\prime}} - m^{\prime} n_{s^{\prime\prime}}\right)}\bm{\sigma}_{i, ss^{\prime\prime}}
    \label{SI equation - general pauli matrix with A matrix}
\end{multline}
which can be further simplified for each Stokes polarization parameter. The $\bm{\sigma}_{1}$ matrix for the Stokes $S_{1}$ polarization parameter contains only a diagonal term, so we obtain:
\begin{multline}
    \tilde{\bm{\sigma}}_{1, ss^{\prime}} = \bm{\sigma}_{1, ss^{\prime}} + \\ +
    \frac{2L}{\pi^{2}}\sum_{m^{\prime \prime}} \frac{\sqrt{m^{\prime \prime} m n_{s^{\prime}} n_{s}} \mathbf{A}_{s^{\prime}s}\tilde{\delta}_{m^{\prime \prime} m}}{\left(\left(m^{\prime \prime}\right)^{2} - m^{2}\right) \left(m^{\prime \prime} n_{s} - m n_{s^{\prime}}\right)}\bm{\sigma}_{1, s^{\prime}s^{\prime}} - \\ 
    + \frac{\sqrt{m^{\prime \prime} m^{\prime} n_{s} n_{s^{\prime}}} \mathbf{A}_{ss^{\prime}}\tilde{\delta}_{m^{\prime \prime} m^{\prime}} }{\left(\left(m^{\prime \prime}\right)^{2} -\left( m^{\prime }\right)^{2}\right) \left(m^{\prime \prime} n_{s^{\prime}} - m^{\prime} n_{s}\right)}\bm{\sigma}_{1, ss}.
\end{multline}
This can be further simplified to the following formula:
\begin{figure*}[]
    \centering
    \includegraphics{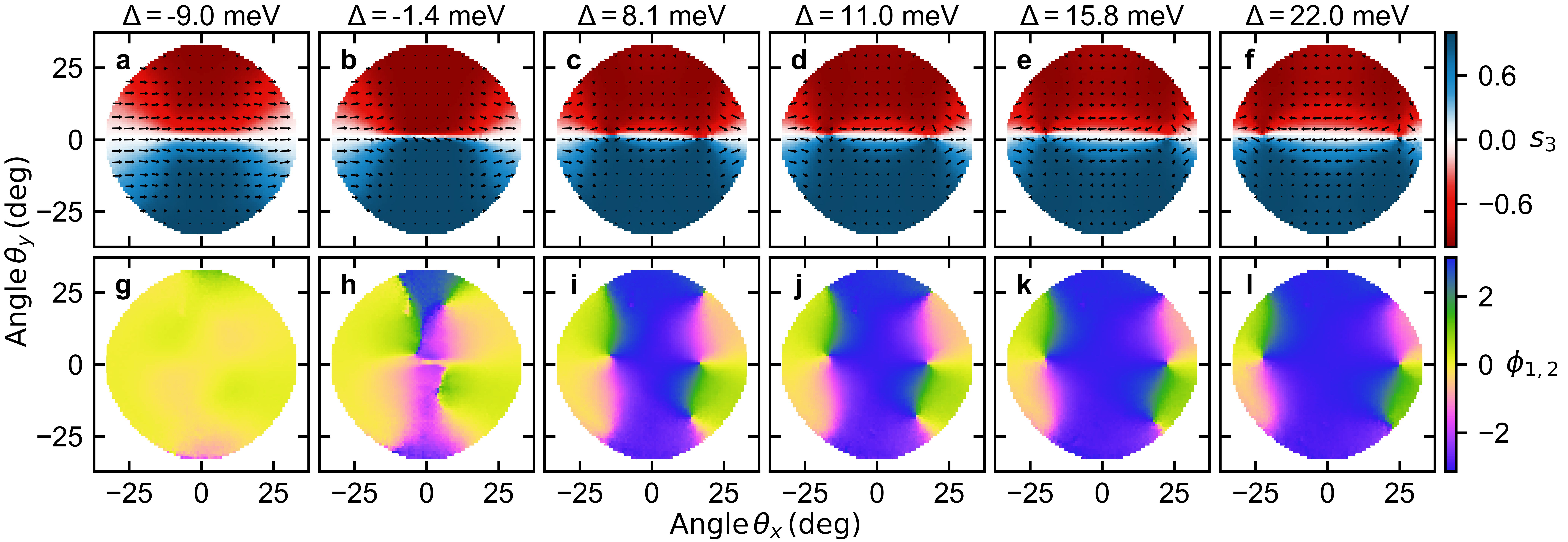}
     \caption{Tunability of the polarization singularities. \textbf{a-f} Lower branch $s_3$ Stokes parameter for as a function of detuning with overlaid $\bm{s}_{\parallel} = \left[s_{1}, s_{2} \right]^{T}$ as black arrows. \textbf{g-i} Corresponding Stokes phase $\phi_{1,2} = \arg \left(s_{1} + is_{2} \right)$ showing changing position of the singularities in reciprocal space.}
    \label{SI - figure: Tunability of C-points for lower branch}
\end{figure*}
\begin{multline}
    \tilde{\bm{\sigma}}_{i, ss^{\prime}} = \tilde{\sigma}_{1, ss^{\prime}} + \frac{2L \sqrt{n_{s} n_{s^{\prime}}} \mathbf{A}_{ss^{\prime}}}{\pi^{2}} \left(g(m, n_{s^{\prime}}) \bm{\sigma}_{1, s^{\prime}s^{\prime}} \right. - \\ + \left. g(m^{\prime}, n_{s}) \bm{\sigma}_{1, ss}\right),
\end{multline}
where the $s$ and $s^{\prime}$ polarizations determine the values of $m$ and $m^{\prime}$, so we set $sm$ and $s^{\prime} m^{\prime}$, which correspond to $X(m + 1)$ or $Y(m)$. The function $g(m, n)$ is a sum over all states present in the cavity and is defined as:
\begin{equation}
    g(m, n) = \sum_{m^{\prime}} \frac{\sqrt{m^{\prime} m} \tilde{\delta}_{m^{\prime} m}}{\left(m^{\prime 2} - m^{2}\right) \left(m^{\prime} n_{s} - m n\right)}
    \label{SI - equation: g function}
\end{equation}
where $n_{s}$ is the refractive index for $s$ polarization, which is the same as for mode number $m$. In this case, the formula for the rotated first Pauli matrix is as follows:
\begin{equation}
    \tilde{\bm{\sigma}}_{1} = 
    \begin{bmatrix}
        1 & -\beta_{1}\\
        \beta_{1} & -1
    \end{bmatrix}
    \label{SI - equation: general first Pauli matrix}
\end{equation}
where $\beta_{1}$ is given by:
\begin{multline}
    \beta_{1} = -i\frac{2L\varepsilon_{xz}\sqrt{n_{X} n_{Y}} k_{y}}{\pi^{2} \varepsilon_{zz}} \left(g\left(m_{Y}, n_{X}\right) \right. +\\ + \left(g \left(m_{X}, n_{Y} \right)\right)
    \label{SI equation: formula for beta1}
\end{multline}
Here, $m_{X}$ and $m_{Y}$ denote the mode numbers for horizontally and vertically polarized modes. The matrix $\tilde{\bm{\sigma}}_{1}$ is Hermitian because $\beta_{1}$ is purely imaginary. In this case, the $S_{1}$ Stokes polarization parameter, according to Eq.~\eqref{SI - equation: polarization parameter general formula} with $\bm{q}_{\pm}$ and $\tilde{\bm{\sigma}}_{1}$ defined in Eqs.~\eqref{SI - equation: eigenvectors} and \eqref{SI - equation: general first Pauli matrix}, is given by:
\begin{multline}
    S_{1, \pm} = \bm{q}^{\dag} \tilde{\bm{\sigma}}_{1} \bm{q} =  \\
    = \begin{bmatrix}
        h_{1} \pm \sqrt{h_{1}^{2} + h_{3}^{2}} \\ 
        i h_{3}
    \end{bmatrix}^{\dag}
    \begin{bmatrix}
        1 & - \beta_{1}\\
        \beta_{1} & 1
    \end{bmatrix}
    \begin{bmatrix}
        h_{1} \pm \sqrt{h_{1}^{2} + h_{3}^{2}} \\ 
        i h_{3}
    \end{bmatrix} = \\
    2 \left(h_{1} - i\beta_{1} h_{3}\right)\left(h_{1} \pm \sqrt{h_{1}^{2} + h_{3}^{2}}\right)
    \label{SI equation: S1 Stokes polarization parameter}
\end{multline}
In this case, since $h_{1} \pm \sqrt{h_{1}^{2} + h_{3}^{2}} \neq 0$, C-points can only occur if $h_{1} - i\beta_{1} h_{3} = 0$.

The same consideration applies to the $S_{2}$ and $S_{3}$ Stokes polarization parameters. In these cases, the Pauli matrices $\bm{\sigma}_{2}$ and $\bm{\sigma}_{3}$ have only anti-diagonal terms (see the comment below in Eq.~\eqref{SI - equation: polarization parameter general formula}), so we simplify the formula for the rotated Pauli matrices given in Eq.~\eqref{SI equation - general pauli matrix with A matrix} to the following formula:
\begin{multline}
    \tilde{\bm{\sigma}}_{i, s s^{\prime}} = \bm{\sigma}_{i, s s^{\prime}} + \\ 
    + \frac{2L}{\pi^{2}} \sum_{m^{\prime \prime}} \frac{\sqrt{m^{\prime \prime} m  n_{\tilde{s}^{\prime}} n_{s}} \mathbf{A}_{\tilde{s}^{\prime} s} \tilde{\delta}_{m^{\prime \prime} m}}{\left(\left( m^{\prime \prime}\right)^{2} - m^{2} \right) \left(m^{\prime \prime} n_{s} - m n_{\tilde{s}^\prime}\right)} \bm{\sigma}_{i, \tilde{s}^{\prime} s^{\prime}}- \\ 
    + \frac{\sqrt{m^{\prime \prime} m^{\prime} n_{\tilde{s}} n_{s^{\prime}}}\mathbf{A}_{\tilde{s} s^{\prime} \tilde{\delta}_{m^{\prime \prime} m^{\prime}}}}{\left( \left( m^{\prime \prime}\right)^{2} - \left(m^{\prime}\right)^{2}\right) \left(m^{\prime \prime} n_{s^{\prime}} - m^{\prime} n_{\tilde{s}}\right)} \bm{\sigma}_{i, s \tilde s}
\end{multline}
where $\tilde{s}$ is defined as:
\begin{equation}
    \tilde{s} = 
    \begin{cases}
    1 &  \mathrm{if} \,s = 2 \\
    2  & \mathrm{if} \,s = 1
    \end{cases}
\end{equation}
The final formula for the rotated Pauli matrices $\tilde{\bm{\sigma}}_{2}$ and $\tilde{\bm{\sigma}}_{3}$ in this case is given by:
\begin{multline}
    \tilde{\sigma}_{i, ss^{\prime}} = \bm{\sigma}_{i, ss^{\prime}} + \\ 
    + \frac{2 L}{\pi^{2}} \left(\sqrt{n_{\tilde{s}_{2}} n_{s_{1}}} \mathbf{A}_{\tilde{s}_{2}s_{1}} g \left(m_{1}, n_{\tilde{s}_{2}} \right) \bm{\sigma}_{i, \tilde{s}_{2} s_{2}} \right. - \\ 
    \left. \sqrt{n_{\tilde{s}_{1}} n_{s_{2}}} \mathbf{A}_{\tilde{s}_{1} s_{2}} g \left(m_{2}, n_{\tilde{s}_{1}} \right) \bm{\sigma}_{i, s_{1} \tilde{s}_{1}} \right)
\end{multline}
where $g\left(m, n\right)$ is defined in Eq.~\eqref{SI - equation: g function}. Finally, the rotated matrices for the $S_{2}$ and $S_{3}$ Stokes polarization parameters are given by:
\begin{subequations}
    \begin{equation}
        \tilde{\bm{\sigma}}_{2} = 
        \begin{bmatrix}
            0 & 1 +\beta_{2} \\
            1 - \beta_{2} & 0
        \end{bmatrix}
    \end{equation}
    \begin{equation}
        \tilde{\bm{\sigma}}_{3} = 
        \begin{bmatrix}
            \beta_{3} g \left(m_{X}, n_{Y} \right) & -i - i\beta_{2} \\
            i - i\beta_{2} & -\beta_{3} g \left(m_{Y}, n_{X} \right)
        \end{bmatrix}
    \end{equation}
\end{subequations}
where:
\begin{subequations}
    \begin{equation}
        \beta_{2} = -i \frac{4L\varepsilon_{xz} n_{X} k_{x}}{\pi^{2} \varepsilon_{zz}} g \left(m_{X}, n_{X} \right)
    \end{equation}
    \begin{equation}
        \beta_{3} = \frac{4L\varepsilon_{xz}\sqrt{n_{X}n_{Y}}k_{y}}{\pi^{2} \varepsilon_{zz}}
    \end{equation}
\end{subequations}
The matrices $\tilde{\bm{\sigma}}_{2}$ and $\tilde{\bm{\sigma}}_{3}$ are Hermitian operators because $\beta_{2}$ is purely imaginary and $\beta_{3}$ is purely real. Using the formulas presented in Eq.~\eqref{SI - equation: eigenvectors} and Eq.~\eqref{SI - equation: polarization parameter general formula}, we obtain the following relations for the $S_{2}$ and $S_{3}$ Stokes polarization parameters:
\begin{figure}[!t]
    \centering
    \includegraphics{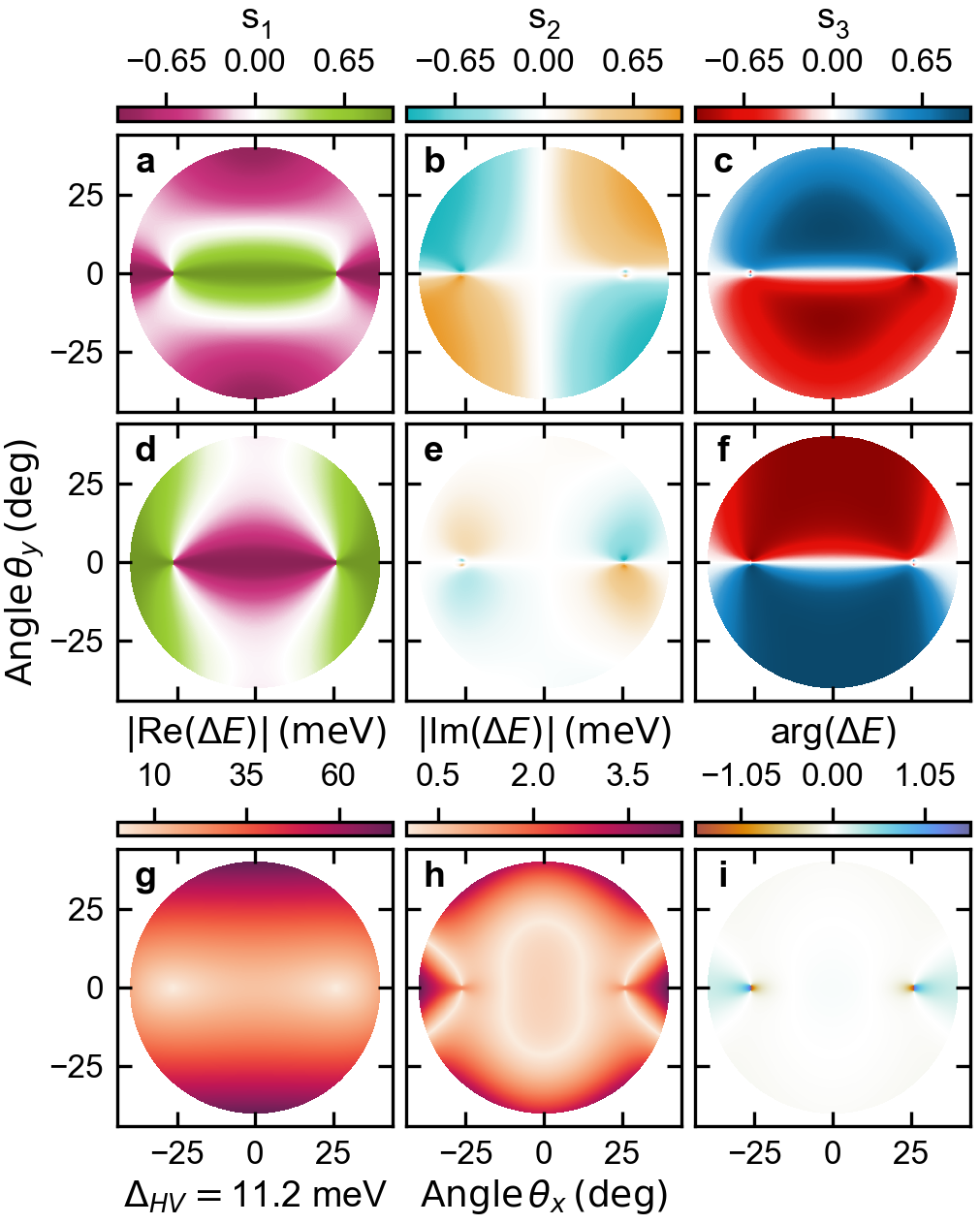}
    \caption{Polarization patterns and difference between energies for two branches for Sample A with 4 pairs of layers in DBRs. \textbf{a}-\textbf{c} and \textbf{d}-\textbf{f} present polarization patterns for upper and lower branches, respectively. \textbf{g} and \textbf{h} presents the absolute value of the real and imaginary part difference between energies for two branches, respectively. \textbf{i}~presents the argument of complex number -- difference between energies for both branches. The $\Delta_{HV}$ denote the difference between energies for two modes for perpendicular incident wave.}
    \label{SI - figure: Polarization patterns for 4 pair of layers in DBRs}
\end{figure}
\begin{subequations}
\begin{multline}
    S_{2, \pm} = \bm{q}^{\dag} \bm{\tilde{\sigma}}_{2} \bm{q} =  \begin{bmatrix}
        h_{1} \pm \sqrt{h_{1}^{2} + h_{3}^{2}}  \\
     i h_{3}
    \end{bmatrix}^{\dag} \cdot \\ 
    \cdot
    \begin{bmatrix}
        0 & 1+ \beta_{2}\\
        1-\beta_{2} & 0
    \end{bmatrix}
    \begin{bmatrix}
        h_{1} \pm \sqrt{h_{1}^{2} + h_{3}^{2}} \\ 
        i h_{3}
    \end{bmatrix} = \\
    2 i\beta_{2} h_{3}\left(h_{1} \pm \sqrt{h_{1}^{2} + h_{3}^{2}}\right)
    \label{SI equation: S2 Stokes polarization parameter}
\end{multline}
\begin{multline}
    S_{3, \pm} = \bm{q}^{\dag} \bm{\tilde{\sigma}}_{3} \bm{q} =  \begin{bmatrix}
        h_{1} \pm \sqrt{h_{1}^{2} + h_{3}^{2}}  \\
     i h_{3}
    \end{bmatrix}^{\dag} \cdot \\ 
    \cdot
        \begin{bmatrix}
            \beta_{3} g \left(m_{X}, n_{Y} \right) & -i - i\beta_{2} \\
            i - i\beta_{2} & -\beta_{3} g \left(m_{Y}, n_{X} \right)
        \end{bmatrix}
    \begin{bmatrix}
        h_{1} \pm \sqrt{h_{1}^{2} + h_{3}^{2}} \\ 
        i h_{3}
    \end{bmatrix} = \\
   \beta_{3} \left[\left(h_{1} \pm \sqrt{h_{1}^{2} + h_{3}^{2}}\right)^{2} g\left(m_{X}, n_{Y}\right) - h_{3}^{2} g \left( m_{Y}, n_{X} \right) \right] + \\ +
   2h_{3}\left(h_{1} \pm \sqrt{h_{1}^{2} + h_{3}^{2}}\right)
\end{multline}
\end{subequations} 
\begin{figure}[!t]
    \centering
    \includegraphics{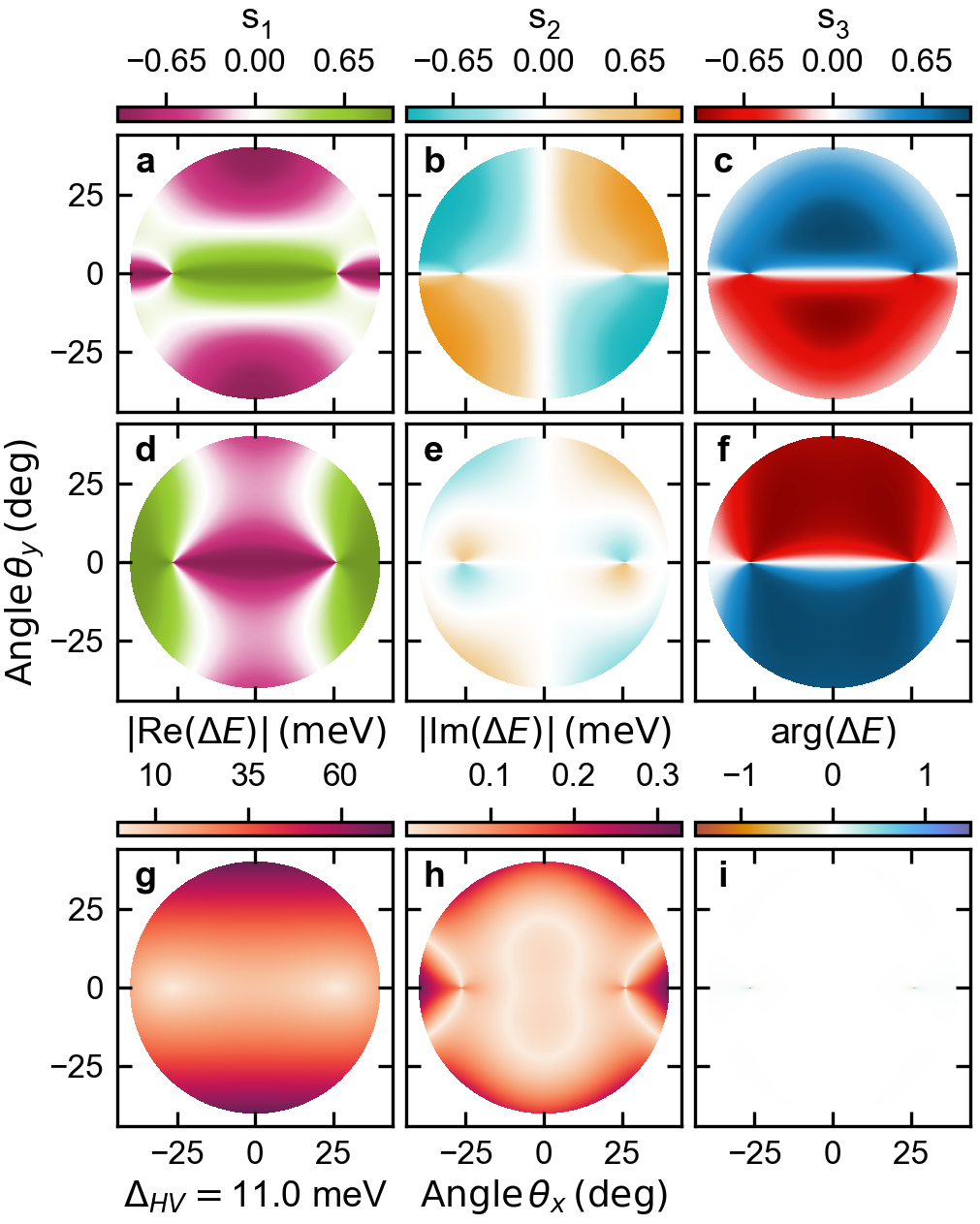}
    \caption{Polarization patterns and difference between energies for two branches for Sample A with 8 pairs of layers in DBRs. \textbf{a}-\textbf{c} and \textbf{d}-\textbf{f} present polarization patterns for upper and lower branches, respectively. \textbf{g} and \textbf{h} presents the absolute value of the real and imaginary part difference between energies for two branches, respectively. \textbf{i}~presents the argument of complex number -- difference between energies for both branches. The $\Delta_{HV}$ denote the difference between energies for two modes for perpendicular incident wave.}
    \label{SI - figure: Polarization patterns for 8 pair of layers in DBRs}
\end{figure}
\newline In the Hermitian limit, $\tilde{\bm{\sigma}}_{2} \neq \bm{\sigma}_{2}$ if $k_{x} \neq 0$, and the $\bm{q}_{\pm}$ vectors are orthogonal. Thus, the $S_{2}$ Stokes polarization parameter is zero at $k_{x} = 0$ or $k_{y} = 0$, but is non-zero in other directions due to the non-zero values of $\beta_{2}$ and $h_{3}$ (see Eq.~\eqref{SI equation: S2 Stokes polarization parameter}). This indicates that the C-points can occur only in the $k_{x} = 0$ direction, since along the $k_{y}=0$ direction we have $h_{3} = 0$, making the $S_{3}$ Stokes polarization parameter zero.

According to the formula presented in Eq.~\eqref{SI equation: S1 Stokes polarization parameter}, the $S_{1}$ Stokes polarization parameter is zero if \mbox{$h_{1} - i \beta_{1} h_{3} = 0$}. Additionally, the $S_{2}$ Stokes polarization parameter is zero at $k_{x} = 0$. Therefore, using the formulas for $h_{1}$ and $h_{3}$ defined in Eq.~\eqref{SI equation: all parameters of Hamiltonian} and $\beta_{1}$ defined in Eq.~\eqref{SI equation: formula for beta1}, we obtain the following relation for the position of the \mbox{C-points} in $k$-space:
\begin{figure}[!t]
    \centering
    \includegraphics{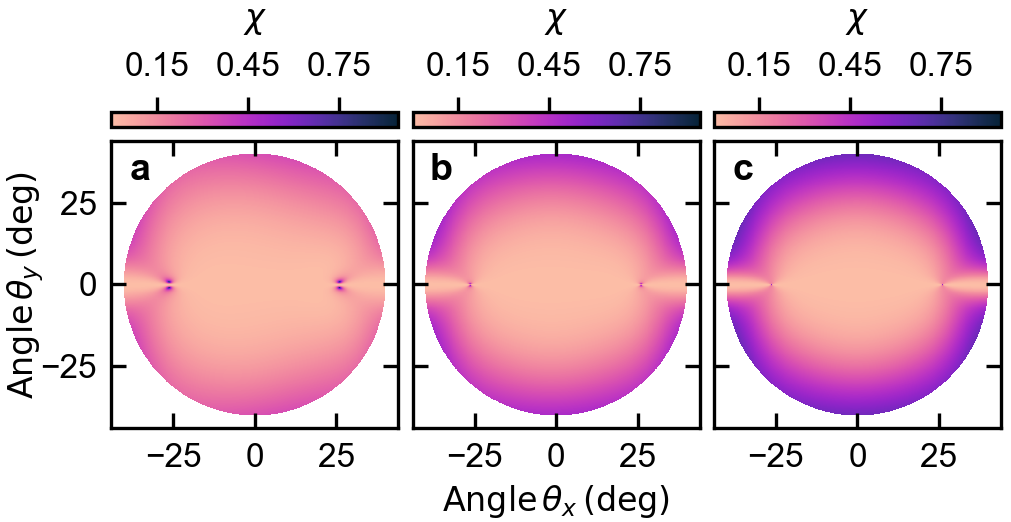}
    \caption{Value of non-othogonality $\chi$ in a function of number of pairs of layers in DBRs for Sample A. \textbf{a} -- 4 pairs, \textbf{b} -- 6 pairs, \textbf{c} -- 8 pairs. The rest parameters are the same as for Sample A.}
    \label{SI - figure: Value of nonorthogonality - sample A}
\end{figure}
\begin{equation}
    k_{y, CP} = \pm\sqrt{\frac{\Delta}{\Sigma \left(m_{Y} \right) \frac{4L c \sqrt{\left(m + 1 \right) m}}{\left(2m + 1\right) \pi^{3}} \left( \frac{\varepsilon_{xz}}{\varepsilon_{zz}} \right)^{2} - \delta_{y}}}
\end{equation}
where $\Sigma\left(m\right) = g\left(m_{Y}, n_{X}\right) + g \left(m_{Y} + 1, n_{Y}\right)$. The sign of $\delta_{y}$ in the Rashba-Dresselhaus coupling depends on the cavity thickness (it is negative for small thickness and positive for a thick cavity). Since $\Sigma\left(m_{Y}\right)$ is always positive, C-points can occur only for positive detuning ($\Delta > 0$).

\subsection{Parameters of simulation}
In this approach, we use the following values for all parameters. The dielectric tensor $\hat{\varepsilon}$ is equal to:
\begin{equation}
    \bm{\varepsilon}\left(\theta \right) = \mathbf{R}_{y} \left(\theta \right) \bm{\varepsilon}_{d} \mathbf{R}_{y}^{T} \left(\theta\right)
\end{equation}
where $\mathbf{R}_{y}\left(\theta\right)$ denotes the standard rotation matrix along $y$ axis, $\bm{\varepsilon}_{d} = \mathrm{diag} \left(n_{e}^{2}, n_o^{2}, n_{o}^{2} \right)$, $n_{o} = 1.5$ and $n_{e} = 1.7$. Cavity thickness is equal to $L = 2120 \,\mathrm{nm}$. The cavity photon lifetime is determined by the value of $\zeta_s$ [see Eq.~\eqref{eq.om0}], which is equal to $\zeta_{X} = 82 \,\mathrm{\mu m^{-1}}$ and $\zeta_{Y} = 78 \,\mathrm{\mu m^{-1}}$ for horizontally and vertically polarized mode, respectively. We performed simulation for $13$-th horizontal and $12$-th vertical mode in a cavity. Using the perturbation theory we take into account the interaction with modes with mode numbers from 5 to 20. The rotation angle of molecule $\theta = 1.02\theta_{r}$ and $\theta = 0.98\theta_{r}$ for positive and negative detuning, respectively, where $\theta_{r}$ denotes the angle for which the $\Delta = 0$. The refractive index outside cavity is equal $n_{a} = 1$. 
\begin{figure}[!t]
    \centering
    \includegraphics{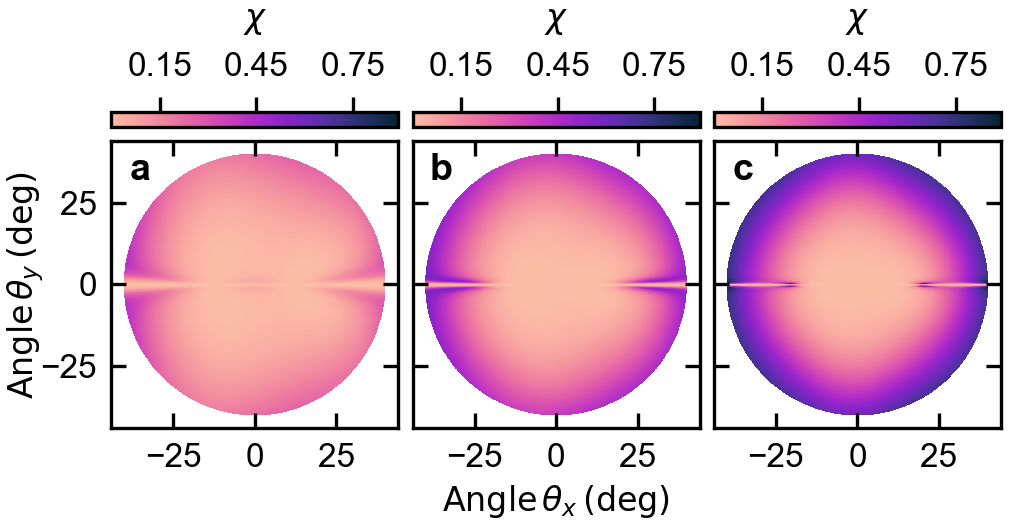}
    \caption{Value of non-othogonality $\chi$ in a function of number of pairs of layers in DBRs for Sample B. \textbf{a} -- 4 pairs, \textbf{b} -- 6 pairs, \textbf{c} -- 8 pairs. The rest parameters are the same as for Sample B.}
    \label{SI - figure: Value of nonorthogonality - sample B}
\end{figure}
\section{Transfer matrix method}
\subsection{General information}
In our approach we used the standard transfer matrix method presented in Berreman's \cite{Berreman_(1972)_J.Opt.Soc.Am.} and Schubert's \cite{ Schubert_(1996)_Phys.Rev.B} papers. In this approach the relation between the two-dimensional vector (in $x - y$ plane, which is the same as the plane of the cavity) of electric $\bm{E} \left(z \right)$ and magnetic $\bm{H} \left(z \right)$ field is given by the following equation:
\begin{multline}
    \partial_{z} 
    \begin{bmatrix}
        \bm{E} \left( z \right) \\
        \bm{H} \left( z \right)
    \end{bmatrix}  = \\ =
    \begin{bmatrix}
        \mathbf{S}_{11} & \mathbf{S}_{12} \\
        \mathbf{S}_{21} & \mathbf{S}_{22} 
    \end{bmatrix}
    \begin{bmatrix}
        \bm{E} \left( z \right) \\ 
        \bm{H} \left( z \right)
    \end{bmatrix} = \mathbf{T} \left( z \right)  \begin{bmatrix}
        \bm{E} \left( z \right) \\ 
        \bm{H} \left( z \right)
    \end{bmatrix}
    \label{equation with matrix S}
\end{multline}
where the matrices $\mathbf{S}_{ij}$ depend on $z$-cooridinate and have following form:
\begin{subequations}
    \label{equation: appendix matrix Sij}
    \begin{equation}
        \mathbf{S}_{11} = 
        -\frac{ik_{0}}{\varepsilon_{zz}}
        \begin{bmatrix}
            \kappa_{x}\varepsilon_{zx} & \kappa_{x}\varepsilon_{zy}\\
            \kappa_{y}\varepsilon_{zx}& 
            \kappa_{y}\varepsilon_{zy}
        \end{bmatrix}
    \end{equation}
    
    \begin{equation}
        \mathbf{S}_{12} = 
        \frac{ik_{0}\eta_{0}}{\varepsilon_{zz}}
        \begin{bmatrix}
           \kappa_{x}\kappa_{y} & \varepsilon_{zz}-\kappa_{x}^{2}\\
            \kappa_{y}^{2}-\varepsilon_{zz} & -\kappa_{x}\kappa_{y}
        \end{bmatrix}
        \label{matrix S_{12}}
    \end{equation}
    
    \begin{equation}
        \mathbf{S}_{21} = 
        \frac{ik_{0}}{\eta_{0}}
        \begin{bmatrix}
            -\tilde{\varepsilon}_{yx}-\kappa_{x}\kappa_{y}& -\tilde{\varepsilon}_{yy}+\kappa_{x}^{2}\\
            \tilde{\varepsilon}_{xx}-\kappa_{y}^{2} & \tilde{\varepsilon}_{xy}+\kappa_{x}\kappa_{y}
        \end{bmatrix}
    \end{equation}
    
    \begin{equation}
        \mathbf{S}_{22} = \frac{ik_{0}}{\varepsilon_{zz}}
        \begin{bmatrix}
            -\kappa_{y}\varepsilon_{yz} & \kappa_{x}\varepsilon_{yz}\\
            \kappa_{y}\varepsilon_{xz}& -\kappa_{x}\varepsilon_{xz}
        \end{bmatrix}.
    \end{equation}
\end{subequations}
\newline Here $\eta_{0} = \sqrt{{\mu_{0}}/{\varepsilon_{0}}}$ denotes the standard vacuum impedance, $\varepsilon_{ij}$ is the element of dielectric tensor, \mbox{$\tilde{\varepsilon}_{ij} = \varepsilon_{ij} - {\varepsilon_{iz}\varepsilon_{zj}} /{\varepsilon_{zz}}$}, $k_{0} = 2\pi/\lambda_{0}$ (wavewvector in vacuum), $\kappa_{x} = n_{a} \sin \left(\theta \right) \cos \left( \varphi \right)$ and $\kappa_{y} = n_{a} \sin \left( \theta \right) \sin \left( \varphi \right)$, where $n_{a}$ is a refractive index outside the structure (in this case it is the refractive index for air, so $n_a = 1$), $\theta$~denotes the angle of incident wave and $\varphi$ is the azimuthal angle of the cavity plane. The transfer matrix for a full cavity is a product of the transfer matrices for all layers presented in the cavity, so:
\begin{subequations}
    \begin{equation}
        \begin{bmatrix}
        \bm{E} \left(z_{0}\right) \\
        \bm{H} \left(z_{0} \right) 
    \end{bmatrix} = \mathbf{T}_{f} 
    \begin{bmatrix}
        \bm{E} \left(z_{N}\right) \\
        \bm{H} \left(z_{N} \right) 
    \end{bmatrix}
    \end{equation}
    \begin{equation}
        \mathbf{T}_{f} = \prod_{i = 1}^{N} \exp \left(\mathbf{T} \left(z_{i}\right) \left(z_{i - 1} - z_{i}\right)\right)
    \end{equation}
\end{subequations}
In standard approach the amplitude of incident \mbox{$\bm{A} = \left[A_{TM}, A_{TE}\right]^{T}$}, reflected $\bm{B} = \left[B_{TM}, B_{TE}\right]^{T}$ and transmitted \mbox{$\bm{C} = \left[C_{TM}, C_{TE}\right]^{T}$} is given in the $TE-TM$ basis, so we introduce the matrices $\mathbf{L}_{a}$ and $\mathbf{L}_{f}$ which transform the amplitude from $TE-TM$ basis to $x-y$ basis. These matrices are equal to:
\begin{widetext}
    \begin{equation}
        \mathbf{L}_{a} = 
        \begin{bmatrix}
            -\cos \left( \theta \right) \cos \left( \varphi \right) & \sin \left( \varphi \right) & \cos \left( \theta \right) \cos \left( \varphi \right) & \sin \left( \varphi \right) \\
            -\cos \left(\theta \right) \sin \left( \varphi \right) & - \cos \left( \varphi \right) & \cos \left( \theta \right) \sin \left( \varphi \right) & - \cos \left( \varphi \right) \\
            \frac{n_{a}}{\eta_{0}} \sin \left(\varphi \right) & \frac{n_{a}}{\eta_{0}} \cos \left( \theta \right) \cos \left( \varphi \right) & \frac{n_{a}}{\eta_{0}} \sin \left( \varphi \right) & - \frac{n_{a}}{\eta_{0}} \cos \left(\theta \right) \cos \left( \varphi \right) \\
            -\frac{n_{a}}{\eta_{0}} \cos \left(\varphi \right) & \frac{n_{a}}{\eta_{0}} \cos \left( \theta \right) \sin \left( \varphi \right) & -\frac{n_{a}}{\eta_{0}} \cos \left( \varphi \right) & - \frac{n_{a}}{\eta_{0}} \cos \left(\theta \right) \sin \left( \varphi \right)
        \end{bmatrix}
    \end{equation}
\end{widetext}
\begin{equation}
    \mathbf{L}_{f} = 
    \begin{bmatrix}
        - \cos \left( \vartheta \right) \cos \left( \varphi \right) & \sin \left( \varphi \right) & 0 & 0 \\
        - \cos \left( \vartheta \right) \sin \left( \varphi \right) & - \cos \left( \varphi \right) & 0 & 0 \\
        \frac{n_{f}}{\eta_{0}} \sin \left( \varphi \right) & \frac{n_{f}}{\eta_{0}} \cos \left( \vartheta \right) \cos \left( \varphi \right) & 0 & 0 \\
        -\frac{n_{f}}{\eta_{0}} \cos \left( \varphi \right) & \frac{n_{f}}{\eta_{0}} \cos \left( \vartheta \right) \sin \left( \varphi \right) & 0 & 0 \\
    \end{bmatrix}
\end{equation}
where $n_{f}$ denotes the refractive index of the isotropic medium at the right side of the sample (air, in our case, so $n_{f} = 1$) and \mbox{$\vartheta = \arcsin \left(n_{a} \sin \left( \theta \right) / n_{f} \right)$}. The final relation between the amplitudes of incident, reflected, and transmitted waves in $TE-TM$ basis is given by following equation:
\begin{equation}
    \label{equation: transfer matrix in TE-TM basis for full system}
    \begin{bmatrix}
        \bm{A} \\ \bm{B} 
    \end{bmatrix} = \mathbf{L}_{a}^{-1} \mathbf{T}_{f} \mathbf{L}_{f} 
    \begin{bmatrix}
        \bm{C} \\ \bm{0}  
    \end{bmatrix} = 
    \begin{bmatrix}
        \mathbf{T}_{11} & \mathbf{T}_{12} \\
        \mathbf{T}_{21} & \mathbf{T}_{22}
    \end{bmatrix}
    \begin{bmatrix}
        \bm{C} \\ \bm{0}  
    \end{bmatrix}
\end{equation}
where $\mathbf{T}_{ij}$ denotes the $2 \times 2$ submatrices of full transfer matrix between amplitude in $TE-TM$ basis. In such a~way (scattering geometry), the amplitudes of reflected and transmitted waves are equal to:
\begin{subequations}
    \begin{equation}
        \bm{B} = \mathbf{T}_{21} \mathbf{T}_{11}^{-1} \mathbf{A}
    \end{equation}
    \begin{equation}
        \bm{C} = \mathbf{T}_{11}^{-1} \bm{A}
    \end{equation}
\end{subequations}
In scattering geometry some effects for example exceptional points (degeneracy points in reciprocal space) cannot be observed in simulation results, so we used another approach, which is called outgoing wave boundary conditions, for which the amplitude of an incident wave is equal to zero. In this approach, the Eq.~\eqref{equation: transfer matrix in TE-TM basis for full system} can be rewritten to the following formula:
\begin{equation}
    \label{equation: transfer matrix in TE/TM basis in outgoing wave boundary condition}
    \begin{bmatrix}
        \mathbf{T}_{11} & \bm{0} \\
        \mathbf{T}_{21} & -\bm{\sigma}_{0}
    \end{bmatrix}
    \begin{bmatrix}
        \bm{C} \\ \bm{B} 
    \end{bmatrix} =
    \begin{bmatrix}
        \bm{0} \\ \bm{0}  
    \end{bmatrix}
\end{equation}
where $\bm{\sigma}_{0}$ denotes the $2\times2$ identity matrix. A nontrivial solution ($\bm{B} \neq \bm{C} \neq \bm{0}$) exists if $\mathrm{det} \left(\mathbf{T}_{11} \right) = 0$, which is satisfied only for complex energy values \mbox{$E = E_{r} - i \Gamma$}. Here, $E_{r}$ and $\Gamma$ represent the resonant energy and the lifetime of the photon in the cavity, respectively. The complex solution was found, by using the standard Newton algorithm for complex root, with the initial value calculated as a quadratic extrapolation based on the three previous values. 

The coefficient $\bm{A}$, $\bm{B}$, and $\bm{C}$ is given in the $TE-TM$ basis and it yields the normalized Stokes polarization parameters in $TE-TM$ basis, which are equal to:
\begin{subequations}   
    \label{SI - equation: Stokes polarization parameters in TE-TM basis}
    \begin{equation}
        \tilde{s}_{0} = \left|v_{TM}\right|^{2} + \left|v_{TE} \right|^{2}
    \end{equation}
    \begin{equation}
        \tilde{s}_{1} = \left|v_{TM}\right|^{2} - \left|v_{TE} \right|^{2}
    \end{equation}
    \begin{equation}
        \tilde{s}_{2} = v_{TM}^{\ast} v_{TE} + v_{TE}^{\ast} v_{TM}
    \end{equation}
    \begin{equation}
        \tilde{s}_{3} = i \left(v_{TE}^{\ast} v_{TM} - v_{TM}^{\ast} v_{TE} \right)
    \end{equation}
\end{subequations}
where $\bm{v}$ denotes the $\bm{A}$, $\bm{B}$ or $\bm{C}$ amplitudes. In the scattering geometry, the incident wave is normalized, so \mbox{$\left|\bm{A}\right| = 1$}. In the outgoing wave geometry, \mbox{$\left|\bm{A}\right| = 0$}, so we normalize the $\bm{B}$ and $\bm{C}$ amplitudes such that \mbox{$\left|\bm{B}\right| = \left|\bm{C}\right| = 1$}. In this experiment, the intensity of light for certain polarization was collected in the $x - y$ basis, so the normalized Stokes polarization parameters in this basis are given by:
\begin{subequations}   
    \label{SI - equation: Stokes polarization parameters in x-y basis}
    \begin{equation}
        s_{0} = \left|u_{x}\right|^{2} + \left|u_{y} \right|^{2}
    \end{equation}
    \begin{equation}
        s_{1} = \left|u_{x}\right|^{2} - \left|u_{y} \right|^{2}
    \end{equation}
    \begin{equation}
        s_{2} = u_{x}^{\ast} u_{y} + u_{y}^{\ast} u_{x}
    \end{equation}
    \begin{equation}
        s_{3} = i \left(u_{y}^{\ast} u_{x} - u_{x}^{\ast} u_{y} \right)
    \end{equation}
\end{subequations}
where $\bm{u}$ denotes the amplitude of incident, reflected or transmitted wave in $x - y$ basis. The relation between the amplitudes in $TE-TM$ basis and $x-y$, which is given by:
\begin{equation}
    \begin{bmatrix}
        u_{x} \\ u_{y} 
    \end{bmatrix} = 
    \begin{bmatrix}
        -\cos \left(\theta_{1}\right) \cos \left( \varphi \right) & \sin \left( \varphi \right) \\
        - \cos \left(\theta_{1} \right) \sin \left(\varphi \right) & - \cos \left(\varphi \right)
    \end{bmatrix} 
    \begin{bmatrix}
        v_{TM} \\ v_{TE}
    \end{bmatrix}
\end{equation}
where $\theta_{1} = \theta$ for positive direction and $\theta_{1} = \pi - \theta$ for negative direction of propagation. Using the relation between these two bases, we find the relation between Stokes polarization patterns in the $TE-TM$ basis Eq.~\eqref{SI - equation: Stokes polarization parameters in TE-TM basis} and $x-y$ basis Eq.~\eqref{SI - equation: Stokes polarization parameters in x-y basis}, which has the following form for $\theta = 0 $ (parallel beam for polarization setup):
\begin{equation}
    \label{SI - equation: transformation of polarization}
    \begin{bmatrix}
        s_{0} \\ s_{1} \\ s_{2} \\ s_{3} 
    \end{bmatrix} =
    \begin{bmatrix}
        1 & 0 & 0 & 0 \\
        0 & \cos \left(2 \varphi \right) & \mp \sin \left(2 \varphi \right) & 0 \\
        0 & \pm \sin \left(2 \varphi \right) & \cos \left(2 \varphi \right) & 0 \\ 
        0 & 0 & 0 & 1
    \end{bmatrix}
    \begin{bmatrix}
        \tilde{s}_{0} \\ \tilde{s}_{1} \\ \tilde{s}_{2} \\ \tilde{s}_{3} 
    \end{bmatrix}
\end{equation}
where the upper sign is for a wave that propagates on the right side (positive $z$) and the lower sign is for the wave that propagates on the left side (negative $z$). 
\subsection{Parameters of the sample}
\subsubsection{Sample A}
\begin{figure}[!t]
    \centering
    \includegraphics{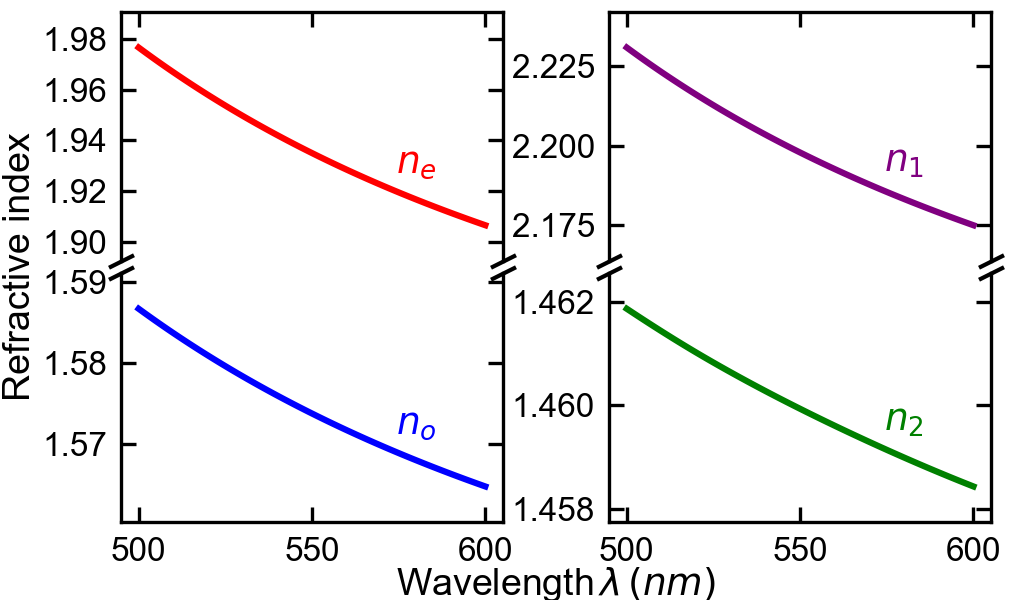}
    \caption{Dispersion relation of refractive indices for liquid crystal used in sample A -- blue and red curves denote the ordinary and extraordinary refractive indices, respectively. Dispersion relation of refractive indices for materials used in DBR -- green $\mathrm{SiO}_{2}$ and purple $\mathrm{TiO}_{2}$.}
    \label{SI - figure: Dispersion relation for materials}
\end{figure}
\begin{figure}[!t]
    \centering
    \includegraphics{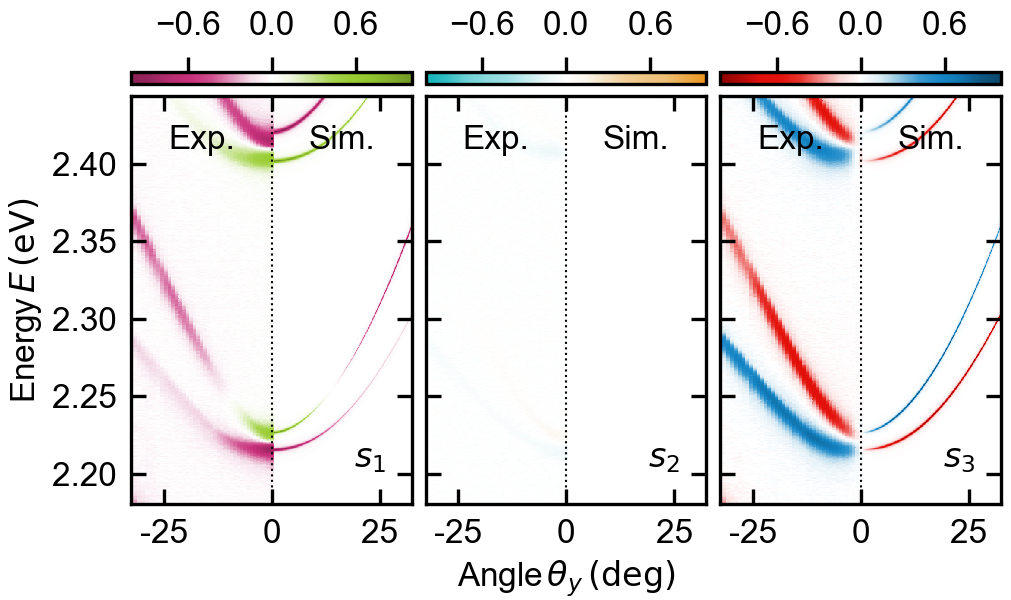}
    \caption{Comparison of the dispersion relation between the experimental data (Exp.) and numerical data (Sim.) for the first sample. The $s_{i}$ denotes the Stokes polarization parameter. The opposite colour for $s_{3}$ parameter is due to the opposite sign of angle for experiment and theory.}
    \label{SI - figure: Dispersion relation for positive detuning}
\end{figure}

Sample A consists of two Distributed Bragg Reflectors (DBRs) centered at 550 nm, made from 6 pairs of $\mathrm{TiO_{2}/SiO_{2}}$ layers, with $\mathrm{SiO_{2}}$ as the top layer. The dispersion relations of refractive index for $\mathrm{TiO}_{2}$ and $\mathrm{SiO_{2}}$ layers are presented in Fig.~\ref{SI - figure: Dispersion relation for materials}. In this calculation, we assumed that the cavity consists of three layers between two DBRs. The first and third layers are the same and isotropic, with thickness 50~nm and refractive index $n = 1.5$. The second layer is a liquid crystal layer with $n_{o}$ and $n_{e}$, which depends on the wavelength, and this dependence is presented in Fig.~\ref{SI - figure: Dispersion relation for materials}. The thickness of this layer is equal to 1163~nm. Dielectric tensor for different angles of liquid crystal molecule is defined as:
\begin{equation}
    \bm{\varepsilon} \left(\theta\right) = 
    \mathbf{R}_{y} \left(\theta \right) \bm{\varepsilon}_{d} \mathbf{R}_{y}^{T} \left( \theta \right)
\end{equation}
where $\mathbf{R}_{y} \left(\theta\right)$ denotes the standard rotation matrix along $y$ axis and $\bm{\varepsilon}_{d} = \mathrm{diag} \left(n_{e}^{2}, n_{o}^{2}, n_{o}^{2} \right)$. The angle of rotation of the molecule is equal to $\theta = 31.4^{\circ}$. The comparison between experimental data collected for this sample and numerical data was presented in Fig.~\ref{SI - figure: Dispersion relation for positive detuning}.
\subsubsection{Sample B}
\begin{figure}[!t]
    \centering
    \includegraphics{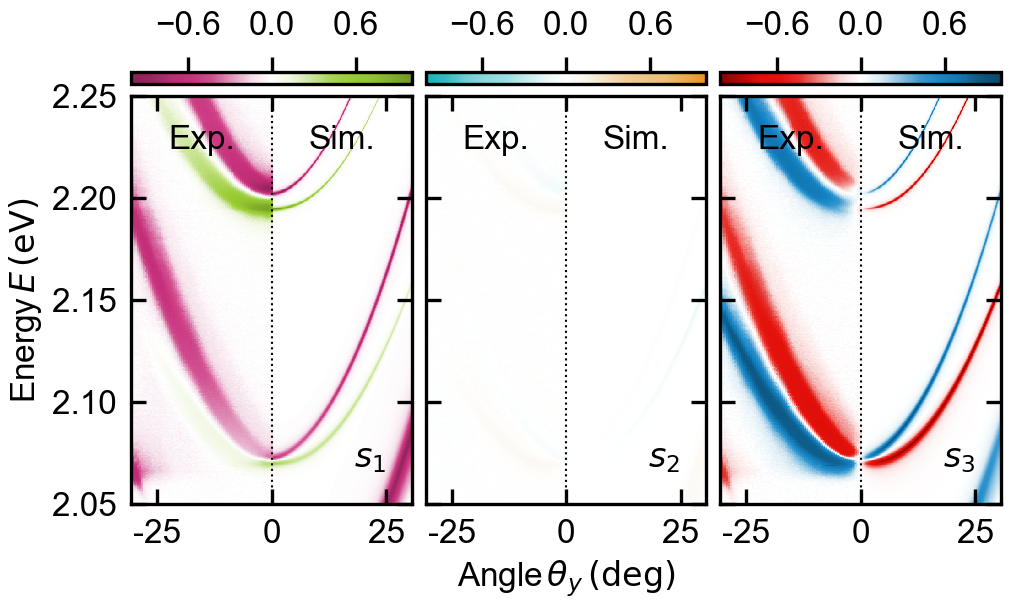}
    \caption{Comparison of the dispersion relation between the experimental data (Exp.) and numerical data (Sim.) for the second sample. The $s_{i}$ denotes the Stokes polarization parameter. The opposite colour for $s_{3}$ parameter is due to the opposite sign of angle for experiment and theory.}
    \label{SI - figure: Dispersion relation for negative detuning}
\end{figure}

Sample B consists of two Distributed Bragg Reflectors (DBRs) centered at 530 nm, made from 6 pairs of $\mathrm{TiO_{2}/SiO_{2}}$ layers, with $\mathrm{SiO_{2}}$ as the top layer. The dispersion relations of refractive index for $\mathrm{TiO}_{2}$ and $\mathrm{SiO_{2}}$ layers are presented in Fig.~\ref{SI - figure: Dispersion relation for materials}. In this calculation, we assumed that the cavity consists of three layers between two DBRs. The first and third layers are the same and isotropic, with thickness 30~nm and refractive index $n = 1.5$. The second layer is a liquid crystal layer with $n_{e} = 1.7114$ and $n_{o} = 1.5190$, which are independent of wavelength. The thickness of this layer is equal to 2385~nm. The angle of rotation of the molecule in the simulation is equal to $\theta = 33.5^\circ$. The comparison between experimental data collected for this sample and numerical data was presented in Fig.~\ref{SI - figure: Dispersion relation for negative detuning}.
\section{Meron structure}
All types of meron structures can be described using the general expression for the Stokes polarization parameter in polar coordinates, which is given by:

\begin{equation}
    \bm{s} = 
    \begin{bmatrix}
        \cos \left( \Phi \left(\varphi\right) \right) \sin \left(\Theta \left(r \right)\right) \\
        \sin \left( \Phi \left(\varphi\right) \right) \sin \left(\Theta \left(r \right)\right) \\
        \cos\left(\Theta \left(r \right)\right)
    \end{bmatrix}
\end{equation}
where $\Phi\left(\varphi\right) = v \varphi + \gamma$, with $v$ representing the vorticity and $\gamma$ representing the helicity, $\Theta \left( r \right)$ is a function specific to each meron structure. The Stokes polarization parameters for Bloch (Fig. 3\textbf{a} in the main text) and Néel (Fig. 3\textbf{b} in the main text) merons are obtained with the following values: $v = 1$, \mbox{$\Theta \left(r\right) = \pi r/2 r_{max}$}, where $r_{max}$ is the maximum radius, $\gamma = \pi/2$ for the Bloch meron, and \mbox{$\gamma = 0$} for the Néel meron. The Stokes polarization parameters for the Node anti-meron (Fig. 3\textbf{c} in the main text) are obtained with the following values: $v = -1$, $\gamma = -\pi$, and $\Theta\left(x, y\right) = \pi/2 - \pi \mathrm{sgn}\left(y\right)\sqrt{x^{2} + y^{2}}/2 r_{max}$, where $\mathrm{sgn}\left(x\right)$ denotes the sign of $x$ and $r_{max} = \sqrt{x_{max}^{2} + y_{max}^{2}}$ and $x_{max} = y_{max}$.

\section{Results for negative detuning}
For negatively detuned H,V cavity modes we expect the C-points to vanish in accordance with our 2-mode Hamiltonian theory Eq.~(2) in main text, discussed around Fig.~1 and 2 in the main text. In order to keep the negatively detuned cavity system as similar as possible in regards to previous results we used another sample for this regime (referred to as Sample B) with a different liquid crystal mixture. The results of angle-resolved tomography for a relatively low negative detuning value of $2\hbar\Delta = -1.7\,\mathrm{meV}$ (with a linewidth of approximately $3\,\mathrm{meV}$) for both upper and lower branches are presented in Figs.~\ref{Figure main - Comparison between experiment and theory - negative detuning}\textbf{a-d} and \ref{Figure main - Comparison between experiment and theory - negative detuning}\textbf{e-h}. The results confirm the absence of C-points in the upper branch, while revealing the presence of two lemon-type $w=1/2$ \mbox{C-points} in the lower branch at high angles $\bm{\theta} = \left[5^{\circ}, 28^{\circ}\right]^{T}$ and $\bm{\theta} = \left[0^{\circ}, -28^{\circ}\right]^{T}$ with a polarization distribution similar to a Bloch meron (see Fig.~\ref{SI - figure: Divergence and Stokes phase for Sample B negative detuning}). Corresponding Berreman and Schubert simulations are shown in Figures~\ref{Figure main - Comparison between experiment and theory - negative detuning}\textbf{i-l} and \ref{Figure main - Comparison between experiment and theory - negative detuning}\textbf{m-p} with good agreement. Comparison between experimental and theoretical dispersion relations for this sample is illustrated in Fig.~\ref{SI - figure: Dispersion relation for negative detuning}.

\bibliography{bibliography}